\documentclass[useAMS,usenatbib,usegraphicx]{mn2e}

\def\plottwo#1#2{\centering \leavevmode
\includegraphics[width=120mm]{#1}
 \hfil \includegraphics[width=120mm]{#2} }

\def\plotthree#1#2#3{\centering \leavevmode
\includegraphics[width=110mm]{#1} \hfil \includegraphics[width=110mm]{#2}
\hfil \includegraphics[width=110mm]{#3} }

\def\smallplottwo#1#2{\centering \leavevmode
\includegraphics[width=85mm]{#1}
 \hfil \includegraphics[width=85mm]{#2} }

\def\smallplot#1{\centering \leavevmode
\includegraphics[width=85mm]{#1} }

\title[Infrared two-colour diagrams for AGB stars using $AKARI$, $MSX$, $IRAS$ and NIR data]
{Infrared two-colour diagrams for AGB stars using $AKARI$, $MSX$, $IRAS$ and NIR data}

\author[Kyung-Won Suh and Young-Joo Kwon]
 {Kyung-Won Suh \thanks{e-mail: kwsuh@chungbuk.ac.kr} and Young-Joo Kwon\\
        Department of Astronomy and Space Science, Chungbuk National University,
        Cheongju-City, 361-763, Republic of Korea}

\begin{document}

\date{Accepted 2011 July 18. Received 2011 July 17; in original form 2011 May 5}

\pagerange{\pageref{firstpage}--\pageref{lastpage}} \pubyear{2011}

\maketitle

\label{firstpage}

\begin{abstract}

Using a revised version of the catalog of AGB stars by Suh \& Kwon
(2009), we present various infrared two-colour diagrams (2CDs) for
3003 O-rich, 1168 C-rich, 362 S-type and 35 silicate carbon stars
in our Galaxy. For each object in the new catalog, we
cross-identify the $AKARI$, $MSX$ and $2MASS$ counterparts by
finding the nearest one from the position information in the $IRAS$
PSC. For the large sample of AGB stars, we present infrared
two-colour diagrams using $IRAS$ (PSC), $AKARI$ (PSC and BSC),
$MSX$ (PSC) and near infrared ($K$ and $L$ bands; including $2MASS$
data at $K_S$ band) data for different classes of AGB stars based
on the chemistry of the dust shell and/or the central star. The
infrared 2CDs of AGB stars can provide useful information about the
structure and evolution of the dust envelopes as well as the
central stars. On the 2CDs, we plot tracks of the theoretical
radiative transfer model results with increasing dust shell optical
depths. Comparing the observations with the theoretical models on
the new 2CDs, we find that the basic model tracks roughly coincide
with the densely populated observed points. Generally, we can
explain the observations of O-rich and C-rich AGB stars on the
various 2CDs with the theoretical models using dust opacity
functions of amorphous silicate, amorphous carbon, SiC and
corundum. For O-rich AGB stars, we find that the models using
corundum as well as silicate can improve the fit with the
observations.

\end{abstract}

\begin{keywords}
stars: AGB and post-AGB - circumstellar matter - infrared: stars -
dust, extinction - radiative transfer.
\end{keywords}

\section{Introduction}

Asymptotic Giant Branch (AGB) stars are generally classified to be
oxygen-rich (M-type) or carbon-rich (C-type) based on the chemistry
of the photosphere and/or the outer envelope (e.g., Busso, Gallino
\& Wasserburg 1999; Herwig 2005). Chan \& Kwok (1990) argued that a
M-type star may become a carbon star when the star goes through C
dredge-up processes and thus the abundance of C is larger than that
of O. S stars are generally regarded as intermediate between M-type
and carbon stars in their properties (e.g., Lloyd Evans \&
Little-Marenin 1999). Only the S stars with Tc (also called
intrinsic S stars) are believed to be actually in the AGB phase
following the evolution sequence M-S-C (Iben \& Renzini 1983;
Jorissen \& Mayor 1988, 1992). However, Chan \& Kwok (1990) and
Guandalini \& Busso (2008) pointed out that the M-S-C evolutionary
sequence is not a certain thing for all the AGB stars.

The two-colour diagram (2CD) of stellar sources in the $Infrared$
$Astronomical$ $Satellite$ ($IRAS$) Point Source Catalog (PSC) has
been useful in characterizing the circumstellar environment of AGB
stars. van der Veen \& Habbing (1988) divided the $IRAS$ 2CD into
eight areas in such a way that each area contains a more or less
homogeneous group. The 2CD statistically distinguishes between
C-rich and O-rich AGB stars. For many studies since the $IRAS$
mission, the $IRAS$ 2CD has been the starting point for colour
selection of samples; not only to select AGB or red giant branch
(RGB) stars, but also to select post-AGB stars, young stellar
objects and starburst galaxies.

The $IRAS$ PSC (version 2.1) contains useful photometric data at
four bands (12, 25, 50 and 100 $\mu$m) for 245,889 sources. $IRAS$
photometric data at four bands as well as near infrared (NIR)
photometric data at $K$ and $L$ bands have been used to make
various infrared 2CDs for AGB stars (e.g., Suh \& Kwon 2009). The
$IRAS$ Low Resolution Spectrograph (LRS; $\lambda$ = 8$-$22 $\mu$m)
data are useful to identify important features of O-rich and C-rich
dust grains in AGB stars (e.g., Kwok, Volk \& Bidelman 1997).

Because the $IRAS$ has a low angular resolution
(0.$\arcmin$75$\times$4.$\arcmin$5-4.$\arcmin$6 pixel size), the
survey regions may suffer from confusion problems. Next infrared
surveys using the $Infrared$ $Space$ $Observatory$ ($ISO$), the
$Midcourse$ $Space$ $Experiment$ ($MSX$), the $Spitzer$ $space$
$telescope$ and the $AKARI$ $space$ $telescope$ ($AKARI$) have
concentrated on the Galactic plane and bulge with much better
angular resolution.

The $MSX$ (Egan et al. 2003) surveyed the Galactic plane as well as
the regions not observed by the $IRAS$ mission with higher
sensitivity and higher spatial resolution (18.3$\arcsec$) in four
mid-infrared broad bands centered at 8.28, 12.13, 14.65 and 21.34
$\mu$m wavelength bands for 441,879 sources.

The $AKARI$ (Murakami et al. 2007) made an all-sky survey with the
infrared camera (IRC) and far infrared surveyor (FIS). We may use
the $AKARI$ PSC data at two bands (9 and 18 $\mu$m) obtained by the
IRC and the bright source catalogue (BSC) data at four bands (65,
90, 140 and 160 $\mu$m ) obtained by the FIS for making meaningful
2CDs.

The two micron all sky survey ($2MASS$; Cutri et al. 2003) used two
highly-automated 1.3-m telescopes equipped with a three-channel
camera capable of observing the sky simultaneously at $J$ (1.25
$\mu$m), $H$ (1.65 $\mu$m) and $K_S$ (2.17 $\mu$m) bands. The PSC
contains accurate positions and fluxes for about 470 million stars
and other unresolved objects.

Nearly all AGB stars can be identified as Long-Period Variables
(LPVs). The general catalog of variable stars (GCVS; Samus et al.
2011) contains the list of LPVs for different variable types. LPVs
in AGB phase are classified according to the amplitude and
regularity of the period in Miras, semi-regulars and irregular
variables. For many pulsating AGB stars, it has been known that the
shapes of the spectral energy distributions (SEDs) vary as well as
the overall luminosity depending on the phase of pulsation. The
shapes of SEDs are affected by the properties of the dust shells as
well as the central stars, depending on the phase of pulsation
(e.g., Suh 2004).

The AGB phase of the LPV is characterized by dusty stellar winds
with high mass-loss rates ($10^{-8} - 10^{-4} M_{\odot}/yr$) (e.g.,
Salpeter 1974; Wachter et al. 2002). Dust envelopes around AGB
stars are believed to be a main source of interstellar dust. The
outflowing envelopes around AGB stars are very suitable places for
massive dust formation (e.g., Kozasa, Hasegawa \& Seki 1984). The
infrared 2CDs of AGB stars can provide useful information about the
structure and evolution of the dust envelopes as well as the
central stars. Comparing the observations with theoretical models,
we may find ways to improve our understanding about the dust
envelopes around AGB stars.

In this paper, we present various infrared 2CDs for O-rich, C-rich,
S-type AGB stars using a revised version of the catalog of AGB
stars by Suh \& Kwon (2009). For each object in the new catalog, we
cross-identify the $AKARI$, $MSX$ and $2MASS$ counterparts by
finding the nearest source from the position information in the
$IRAS$ PSC. For the large sample of AGB stars, we present various
infrared 2CDs using $IRAS$ (PSC), $AKARI$ (PSC and BSC), $MSX$
(PSC) and NIR ($K$ and $L$ bands; including $2MASS$ data at $K_S$
band) data. In this paper, we are concerned about 2CDs for
different classes of AGB stars based on the chemistry of the dust
shell and/or the central star. On the 2CDs, we plot tracks of the
theoretical radiative transfer model results with increasing dust
shell optical depths. Comparing the observations with the
theoretical tracks, we discuss the meaning of the infrared 2CDs.

\section{Sample stars}

For the sample of AGB stars, we use a revised version of the
catalog by Suh \& Kwon (2009). They presented a catalog of AGB
stars in our Galaxy from the sources listed in the $IRAS$ PSC
compiling the lists of previous works with verifying processes;
their catalog was made of 2193 O-rich stars, 1167 C-rich stars, 287
S stars and 36 silicate carbon stars. In this paper, we add SiO
maser sources to the list of O-rich AGB stars and more sources for
C-rich stars and S stars from new references and make some
corrections to the previous catalog. For a general description of
the identifying and verifying processes, refer to Suh \& Kwon
(2009).

\subsection{O-rich stars}

O-rich AGB stars typically show the conspicuous 10 $\mu$m and 18
$\mu$m features in emission or absorption. They suggest the
presence of silicate dust grains in the outer envelopes around them
(e.g., Suh 1999). Low mass-loss rate O-rich AGB (LMOA) stars with
thin dust envelopes show the 10 $\mu$m and 18 $\mu$m emission
features. High mass-loss rate O-rich AGB (HMOA) stars with thick
dust envelopes show the absorbing features at the same wavelengths.

OH maser observations identified many OH/IR stars. Suh \& Kwon
(2009) listed 1533 sources from 14 papers as O-rich AGB stars.
However, methanol maser sources at 6.7 GHz are only associated with
massive star formation (Minier et al. 2003). More than 550 methanol
maser sources have been detected (Pestalozzi, Minier \& Booth
2005). Using the catalog by Pestalozzi et al. (2005), we have
excluded 18 methanol maser sources from the OH maser list in Suh \&
Kwon (2009).

Many of the SiO maser sources are O-rich AGB stars. The catalog of
Suh \& Kwon (2009) did not consider the SiO maser sources. By
compiling the sources listed in 17 related papers, we have added
815 SiO maser sources which are identified to be O-rich AGB stars
(see Table 1).

655 sources detected by other methods including photometric and
spectrometric methods (molecular emission or spectral types) are
compiled from 10 related papers. See Suh \& Kwon (2009) for a
detailed description.

Table 1 shows the revised list of O-rich AGB stars which contains
3003 sources. Compared to Suh \& Kwon (2009), the number has been
increased by 810. For each object, we have cross-identified the
$AKARI$ and $MSX$ sources as we describe in Section 3.

\begin{table*}
\caption{Sample of O-rich AGB stars. \label{tbl-1}}
\begin{flushleft}
\begin{tabular}{lllllll}
\hline  \hline
Detecting Methods &References  &   Original    &   Excluded    &   Selected    &   Duplicate   &   Remaining   \\
\hline
OH maser & Eder, Lewis \& Terzian (1988)   &   182 &   7   &   175 &   0   &   175 \\
 & Lewis, Eder \& Terzian (1990)  &   86  &   1   &   85  &   0   &   85  \\
 & Chengalur et al. (1993)  &   132 &   7   &   125 &   3   &   122 \\
 & Lewis (1994)   &   56  &   5   &   51  &   17  &   34  \\
 & Sivagnanam et al. (1990)   &   36  &   0   &   36  &   6   &   30  \\
 & Le Squeren et al. (1992)   &   115 &   7   &   108 &   36  &   72  \\
 & David et al. (1993)    &   141 &   2   &   139 &   25  &   114 \\
 & Sevenster et al. (1997a)   &   307 &   108 &   199 &   31  &   168 \\
 & Sevenster et al. (1997b)   &   202 &   53  &   149 &   2   &   147 \\
 & Sevenster et al. (2001)    &   286 &   81  &   205 &   67  &   138 \\
 & te Lintel Hekkert et al. (1989)    &   442 &   186 &   256 &   169 &   87  \\
 & Nyman, Hall \& Le Bertre (1993)    &   44  &   5   &   39  &   37  &   2   \\
 & Lepine et al. (1995)   &   405 &   41  &   364 &   223 &   141 \\
 & Chen et al. (2001) &   1065    &   110 &   955 &   737 &   218 \\
\hline
SiO maser & Engels \& Heske (1989)    &   180 &   23  &   157 &   67  &   90  \\
 & Hall et al. (1990a)    &   31  &   0   &   31  &   10  &   21  \\
 & Hall et al. (1990b)    &   29  &   0   &   29  &   15  &   14  \\
 & Jewell et al. (1991)   &   47  &   5   &   42  &   42  &   0   \\
 & Nyman et al. (1993)    &   21  &   3   &   18  &   17  &   1   \\
 & Izumiura et al. (1994) &   50  &   0   &   50  &   7   &   43  \\
 & Izumiura et al. (1995a)    &   61  &   1   &   60  &   3   &   57  \\
 & Izumiura et al. (1995b)    &   77  &   0   &   77  &   20  &   57  \\
 & Jiang et al. (1996)    &   52  &   1   &   51  &   10  &   41  \\
 & Izumiura et al. (1999) &   63  &   1   &   62  &   10  &   52  \\
 & Deguchi et al. (2000a) &   86  &   0   &   86  &   7   &   79  \\
 & Deguchi et al. (2000b) &   122 &   0   &   122 &   18  &   104 \\
 & Ita et al. (2001)  &   26  &   0   &   26  &   10  &   16  \\
 & Nakashima \& Deguchi (2003a)   &   134 &   0   &   134 &   87  &   47  \\
 & Nakashima \& Deguchi (2003b)   &   43  &   0   &   43  &   37  &   6   \\
 & Deguchi et al. (2004) &   254 &   0   &   254 &   73  &   181 \\
 & Deguchi et al. (2007)  &   119 &   103 &   16  &   10  &   6   \\
\hline
Others & Rowan-Robinson et al. (1986) &   95  &   5   &   90  &   54  &   36  \\
 & Epchtein et al. (1990) &   29  &   6   &   23  &   11  &   12  \\
 & Blommaert, van der Veen \& Habing (1993)    &   15  &   1   &   14  &   11  &   3   \\
 & Kastner et al. (1993)  &   19  &   1   &   18  &   9   &   9   \\
 & Loup et al. (1993) &   184 &   12  &   172 &   127 &   45  \\
 & Whitelock et al. (1994)    &   58  &   0   &   58  &   30  &   28  \\
 & Guglielmo et al. (1997)    &   16  &   0   &   16  &   3   &   13  \\
 & Guglielmo et al. (1998)    &   23  &   2   &   21  &   3   &   18  \\
 & Le Bertre et al. (2003)    &   563 &   31  &   532 &   58  &   474 \\
 & Jimenez-Esteban et al. (2006)  &   59  &   0   &   59  &   42  &   17  \\
\hline
Total number    &  &    5908    &   803 &   5105    &   2102    &   3003    \\
\hline
$AKARI$-PSC &     &       &       &       &       &   2356    \\
$AKARI$-BSC &     &       &       &       &       &   1548    \\
\hline
$MSX$-PSC &     &       &       &       &       &   1966  \\
\hline  \hline
\end{tabular}
\end{flushleft}
\end{table*}

\subsection{C-rich stars}

The main components of dust in the envelopes around carbon stars
are believed to be featureless amorphous carbon (AMC) grains and
SiC grains which produce the 11.3 $\mu$m emission feature (e.g.,
Suh 2000). The carbon stars showing the 11.3 $\mu$m emission
feature belong to $IRAS$ LRS class C.

For C-rich AGB stars, we use the same catalog of Suh \& Kwon (2009)
except that one reference (Loup et al. 1993) is added and six
objects are excluded in this work. Because $IRAS$ used a large
aperture, the flux from an $IRAS$ PSC source may contain blended
fluxes from multiple objects. Two objects ($IRAS$ 19075+0921 and
22306+5918) which show blended fluxes and four planetary nebulae
(Kohoutek 2001; $IRAS$ 04395+3601, 05251-1244, 21282+5050 and
21306+4422) are excluded.

Table 2 lists the 1168 C-rich AGB stars as identified by various
authors and verified in this paper and Suh \& Kwon (2009). For each
object, we have cross-identified the $AKARI$ and $MSX$ counterparts
(see Section 3).

\begin{table*}
\caption{Sample of C-rich AGB stars. \label{tbl-2}}
\begin{flushleft}
\begin{tabular}{llllll}
\hline \hline
References  &   Original    &   Excluded    &   Selected    &   Duplicate   &   Remaining   \\
\hline
Rowan-Robinson et al. (1986)$^a$    &   40  &   0  &  41$^a$  &   0   &   41  \\
Chan \& Kwok (1990)   &   145 &   0   &   145 &   22  &   123 \\
Epchtein et al. (1990)    &   216 &   7   &   209 &   60  &   149 \\
Egan \& Leung (1991)  &   125 &   1   &   124 &   78  &   46  \\
Volk, Kwok \& Langill (1992)    &   32  &   0   &   32  &   9   &   23  \\
Chan (1993)   &   106 &   0   &   106 &   55  &   51  \\
Groenewegen, de Jong \& Baas (1993) &   25  &   0   &   25  &   14  &   11  \\
Guglielmo et al. (1993)   &   106 &   5   &   101 &   0   &   101 \\
Kastner et al. (1993) &   18  &   0   &   18  &   9   &   9   \\
Volk, Kwok \& Woodsworth (1993)   &   17  &   2   &   15  &   13  &   2   \\
* Loup et al. 1993 &   205  &   16   &   189  &   160  &   29   \\
Groenewegen, de Jong \& Geballe (1994) &   16  &   0   &   16  &   14  &   2   \\
Lorenz-Martins \& Lefevre (1994)  &   32  &   0   &   32  &   30  &   2   \\
Whitelock et al. (1994)   &   3   &   0   &   3   &   3   &   0   \\
Groenewegen (1995)    &   21  &   0   &   21  &   21  &   0   \\
Groenewegen, van den Hoek \& de Jong (1995) &   21  &   0   &   21  &   10  &   11  \\
Guglielmo et al. (1997)   &   25  &   0   &   25  &   0   &   25  \\
Kwok et al. (1997)    &   715 &   2   &   713 &   402 &   311 \\
Guglielmo et al. (1998)   &   27  &   0   &   27  &   2   &   25  \\
Groenwegen et al. (2002)  &   252 &   14  &   238 &   208 &   30  \\
Le Bertre et al. (2005)   &   143 &   4   &   139 &   38  &   101 \\
Guandalini et al. (2006)  &   270 &   20  &   250 &   241 &   9  \\
Menzies, Feast \& Whitelock (2006) &   177 &   2   &   175 &   146 &   29  \\
Whitelock et al. (2006)   &   257 &   35  &   222 &   180 &   42  \\
Chen \& Shan (2008)   &   348 &   0   &   348 &   348 &   0   \\
\hline
Total number  &   2970    &   111   &   2860    &   1692    &   1168    \\
\hline
$AKARI$-PSC   &       &       &       &       &   1012    \\
$AKARI$-BSC   &       &       &       &       &   675 \\
\hline
$MSX$-PSC   &       &       &       &       &   687    \\
\hline \hline
\scriptsize $^a$: One object is added to the list.
\end{tabular}
\end{flushleft}
\end{table*}

\subsection{S stars and silicate carbon stars}

For the sample of S stars in AGB phase, we use the list of objects
which are classified to be intrinsic S stars by Van Eck et al.
(2000), Yang et al. (2006) and Guandalini \& Busso (2008). Table 3
shows the revised list of S stars which contains 362 sources.
Compared to Suh \& Kwon (2009), the number has been increased by
75. Silicate carbon stars are the carbon stars with silicate dust
features. For the sample of silicate carbon stars in AGB phase, we
use the same list as presented in Suh \& Kwon (2009) except that
one silicate carbon star ($IRAS$ 04496-6958) is excluded because it
is not in our Galaxy. The list of S stars and silicate carbon stars
is shown in Table 3.

\begin{table*}
\caption{S stars and Silicate carbon stars. \label{tbl-3}}
\begin{flushleft}
\begin{tabular}{lllllll}
\hline \hline
References  &   Original    &   Excluded    &   Selected    &   Duplicate   &   Remaining&     Note \\
\hline
Van Eck et al. (2000)      &  133 &  11 &  122 &   0 &  122 &   S stars\\
Yang et al.(2006)          &  286 &   2 &  285 &  52 &  233 &   S stars\\
Guandalini \& Busso (2008) &  187 &   2 &  185 & 178 &    7 &   S stars\\
\hline
Total number               &  606 &  15 &  592 & 230 &  362 &   S stars\\
\hline
$AKARI$-PSC                &      &     &      &     &  336 & \\
$AKARI$-BSC                &      &     &      &     &   86 & \\
\hline
$MSX$-PSC                  &      &     &      &     &  128 & \\
\hline \hline
Kwok \& Chan(1993)         &   15 &     &      &  0  &   15  &   silicate C \\
Chen, Wang \& Wang (1999) &   22 &     &      &  15 &    7  &   silicate C \\
Jiang, Szczerba \& Deguchi (2000)
                           &   1  &     &      &  0  &    1  &   silicate C \\
Molster et al. (2001)      &   1  &     &      &  0  &    1  &   silicate C \\
Chen \& Wang (2001)        &   1  &     &      &  0  &    1  &   silicate C \\
Chen \& Zhang (2006)       &   9  &     &      &  0  &    9  &   silicate C \\
Boboltz et al. (2007)      &   1  &     &      &  0  &    1  &   silicate C \\
\hline
Total number               &   50 &     &      & 15  &   35  &  silicate C  \\
\hline
$AKARI$-PSC                &      &     &      &     &   31 & \\
$AKARI$-BSC                &      &     &      &     &   18 & \\
\hline
$MSX$-PSC                  &      &     &      &     &   22 & \\
\hline \hline
\end{tabular}
\end{flushleft}
\end{table*}

\subsection{NIR data}

For the AGB stars listed in our new catalog, we have tried to
collect all the available NIR data at $K$ and $L$ bands obtained
from ground based observations. We have made revisions to the list
of NIR data in Suh \& Kwon (2009) by verifying the position
information and added $2MASS$ data for many objects.

We cross-identify the $2MASS$ source by finding the nearest one in
the position within 10$\arcsec$ by using the position information
in version 2.1 of the $IRAS$ PSC. We have found the $2MASS$
counterparts for 2840 O-rich stars, 1090 C-rich stars, 356 S stars
and 32 silicate carbon stars. We use only the good quality $2MASS$
NIR data at $K_S$ band for this paper.

Variations of the $K$ band (2.2 $\mu$m) are designated as $K_S$
(2.0-2.3 $\mu$m; 2.17 $\mu$m for $2MASS$). For analyzing the large
sample of data in a single 2CD, we need to ignore the slight
variations of the $K$ and $L$ bands depending on the observation
systems. In this paper, we assume the same band wavelengths for $K$
(2.2 $\mu$m) and $L$ (3.5 $\mu$m) bands for all variations. There
could be minor errors for this assumption.

Table 4 lists the numbers of collected NIR data at $K$ and $L$
bands for the sample AGB stars. Some stars are observed more than
one time.

\begin{table*}
\caption{NIR observations of AGB stars. \label{tbl-4}}
\begin{flushleft}
\begin{tabular}{llllllll}
\hline \hline
  & \multicolumn{2}{l}{O-rich stars}  & \multicolumn{2}{l}{C-rich stars} & S stars & \multicolumn{2}{l}{Silicate carbon stars}\\
\vspace{1mm}
References &   $K$   &   $L$   &   $K$   &   $L$   &   $K$   &   $K$   &   $L$   \\
\hline
Neugebauer et al. (1969)  &       &       &       &       &       &   4   &   4   \\
Epchtein et al. (1990)    &   20  &   20  &   217 &   217 &       &   3   &   3   \\
Lloyd Evans (1990)        &       &       &       &       &       &   3   &   3   \\
Noguchi et al. (1990)     &       &       &       &       &       &   3   &   3   \\
Blommaert et al. (1993)   &   9   &   9   &       &       &       &       &       \\
Groenewegen et al. (1993) &       &       &   25  &   25  &       &       &       \\
Guglielmo et al. (1993)   &       &       &   101 &   101 &       &       &       \\
Kastner et al. (1993)     &   18  &   17  &   15  &   14  &       &       &       \\
Nyman et al. (1993)       &   35  &   35  &       &       &       &       &       \\
Whitelock et al. (1994)   &   58  &   58  &   3   &   3   &       &       &       \\
Xiong et al. (1994)       &   56  &   19  &       &       &       &       &       \\
Lepine et al. (1995)      &   358 &   332 &       &       &       &   1   &   1   \\
Noguchi et al. (1995)     &       &       &       &       &       &   5   &   5   \\
Guglielmo et al. (1997)   &   16  &   16  &   39  &   39  &       &       &       \\
Guglielmo et al. (1998)   &   21  &   21  &   27  &   27  &       &       &       \\
Deguchi et al. (2002)     &   111 &   0   &       &       &       &       &       \\
Le Bertre et al. (2003)   &   535 &   535 &       &       &       &       &       \\
Le Bertre et al. (2005)   &       &       &   139 &   139 &       &   1   &   1   \\
Jimenez-Esteban et al. (2005)
                          &   287 &   0   &       &       &       &       &       \\
Jimenez-Esteban et al. (2006)
                          &   59  &   0   &       &       &       &       &       \\
Whitelock et al. (2006)   &       &       &   222 &   219 &       &   2   &   2   \\
Izumiura et al. (2008)    &       &       &       &       &       &   2   &   2   \\
\hline
$2MASS$ ($K_S$; good quality)    & 1743  &       & 577   &       &  311  &   15  &       \\
\hline
Total number$^a$   &    3326 (2307)    &  1062 (967)    &   1365 (877)   &  784 (658)   &   311(311)  &  39(23)   &  24(15)  \\
$AKARI$-PSC    &    1853    &   766    &   761   &  561   &   297  &  21   &  15  \\
\hline \hline
\scriptsize $^a$: The number in a parenthesis means the number of
objects.
\end{tabular}
\end{flushleft}
\end{table*}

\section{$AKARI$ and $MSX$ data}

By using the position information in version 2.1 of the $IRAS$ PSC,
we cross-identify the $AKARI$ and $MSX$ counterparts by finding the
nearest source for each object in the new catalog of AGB stars.

\subsection{$AKARI$ data}

\begin{figure*}
\smallplottwo{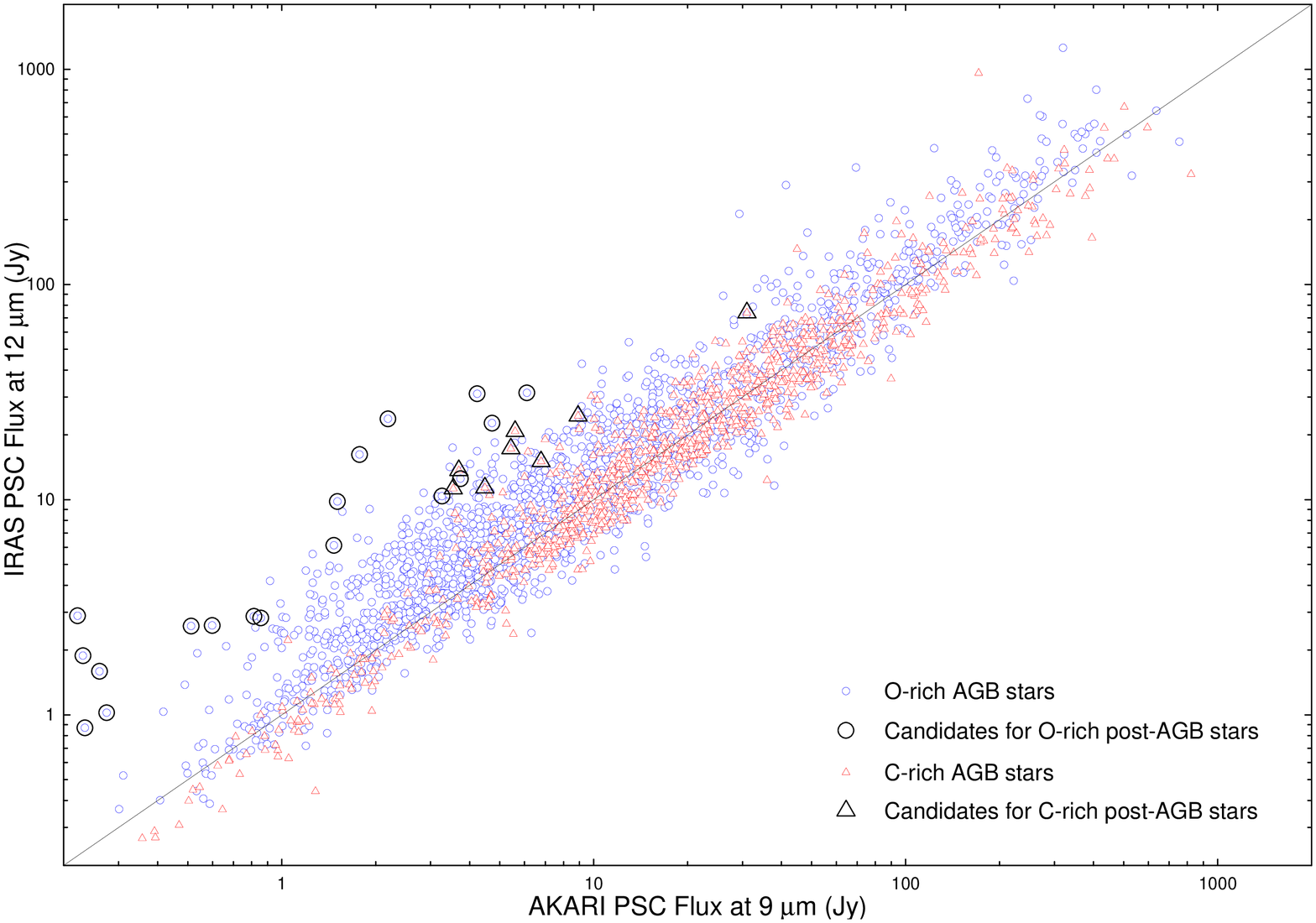}{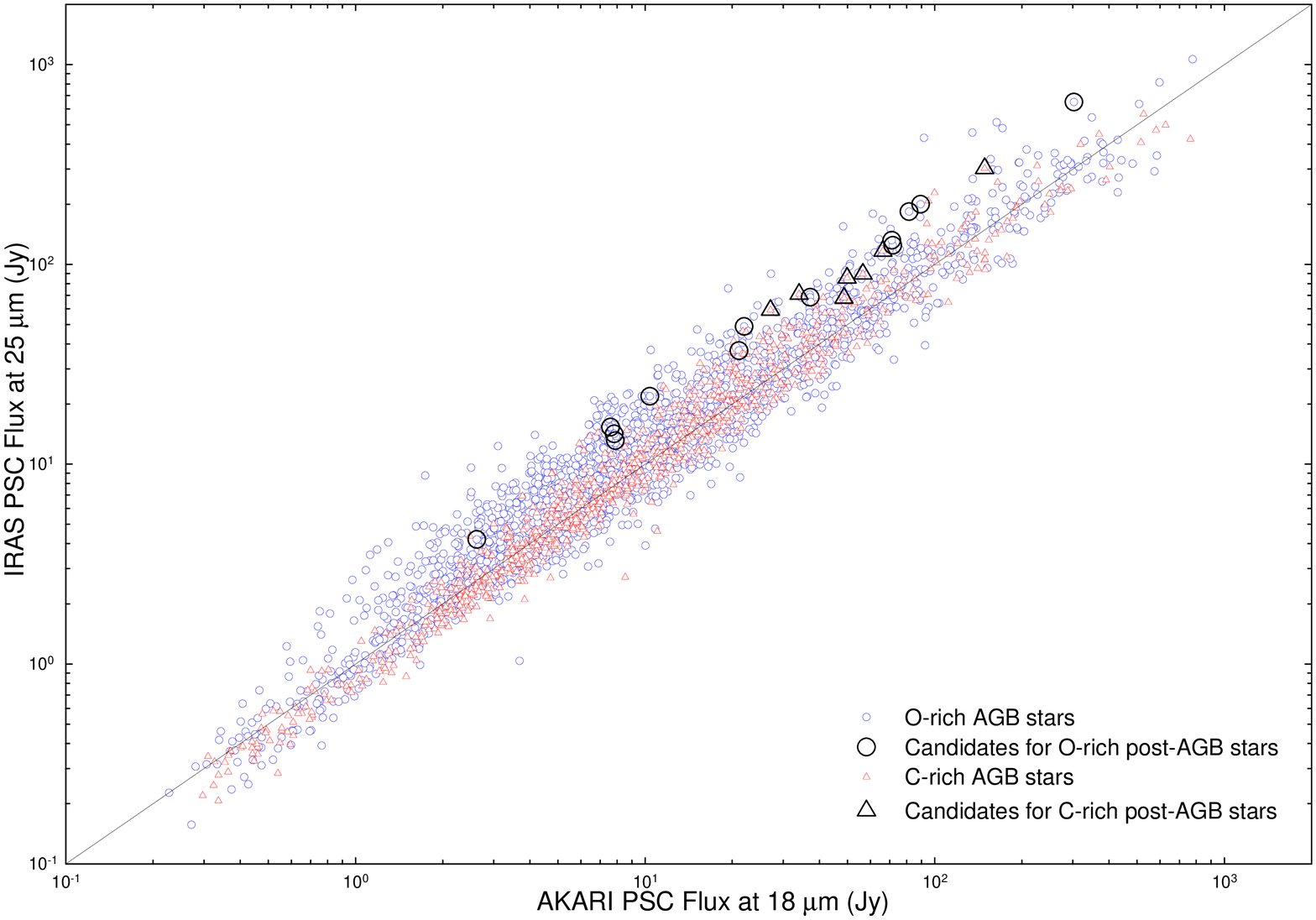} \caption{Comparison of $IRAS$ and
$AKARI$ PSC data for AGB stars.}
\end{figure*}

\begin{figure*}
\smallplottwo{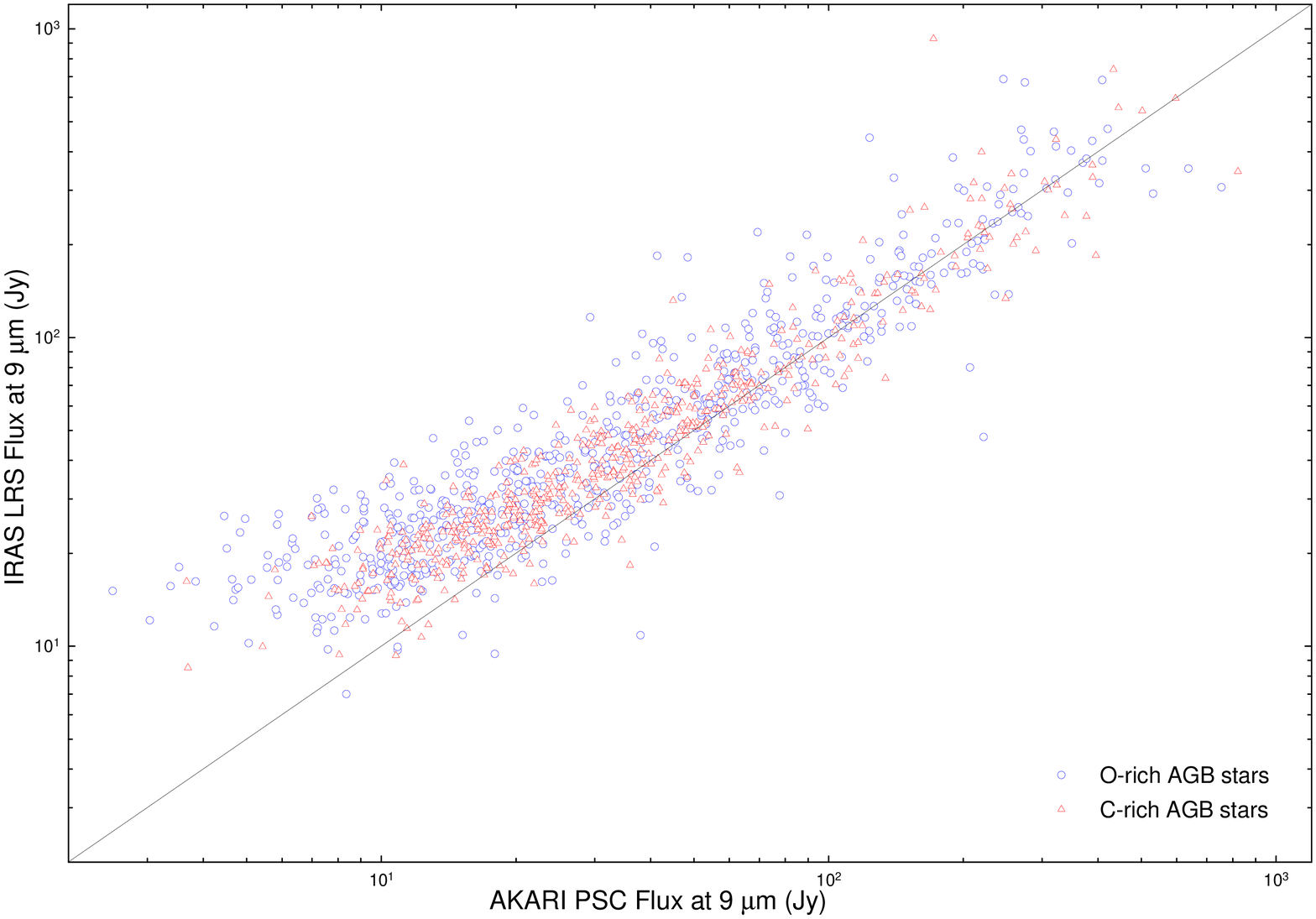}{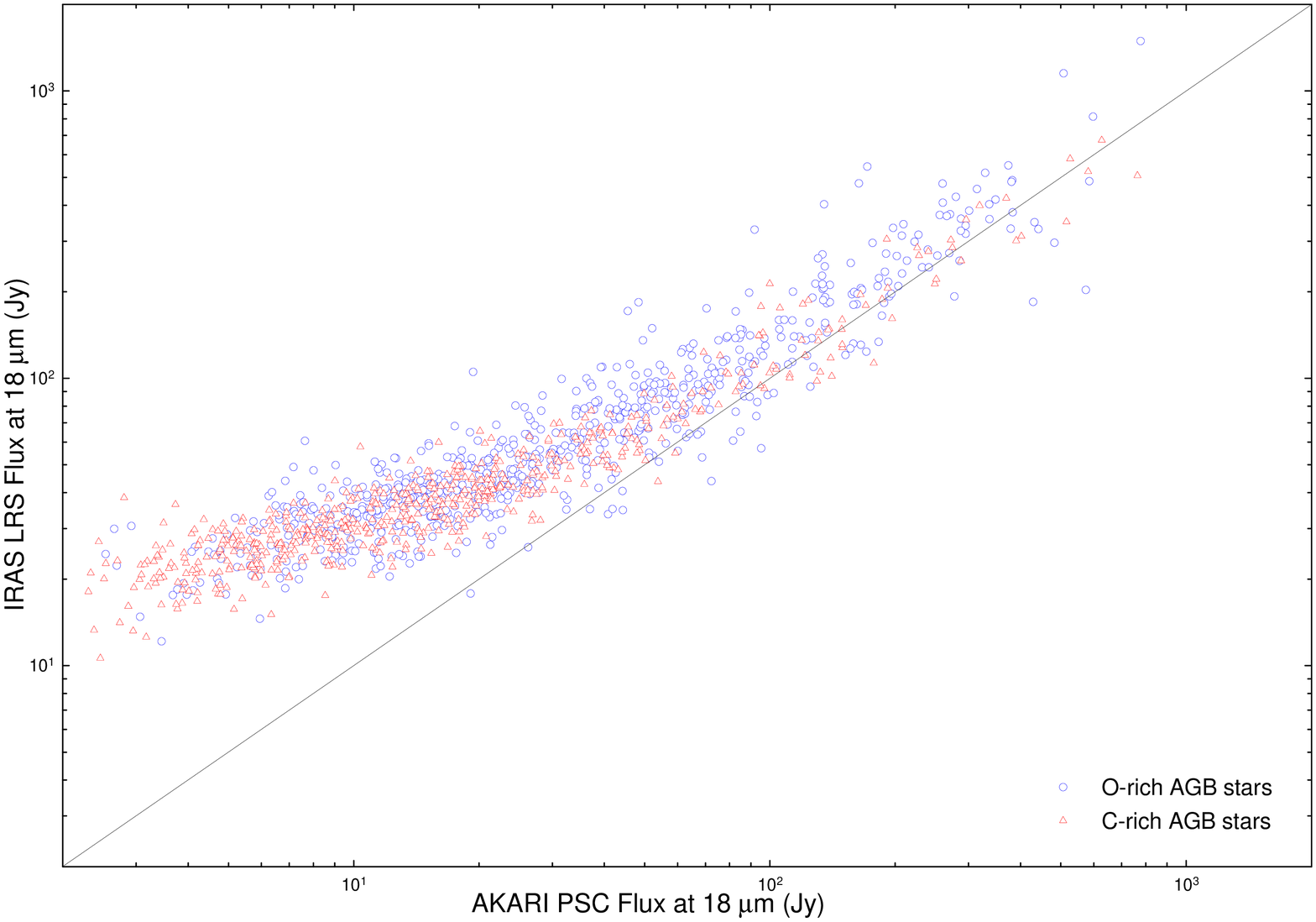} \caption{Comparison of $IRAS$ LRS and
$AKARI$ PSC data for AGB stars.}
\end{figure*}

For $AKARI$ PSC, we find the closest counterpart in the position
within 10$\arcsec$. For $AKARI$ BSC, we find the closest source in
the position within 1$\arcmin$.

We have cross-identified the $AKARI$ PSC counterparts for 2356
O-rich stars (detection rate: 78.5 \%), 1012 C-rich stars (86.6
\%), 336 S stars (92.8 \%) and 31 silicate carbon stars (88.6 \%)
(see Tables 1 through 3). We have cross-identified the $AKARI$ BSC
counterparts for 1548 O-rich stars (51.5 \%), 675 C-rich stars
(57.8 \%), 86 S stars (23.8 \%) and 18 silicate carbon stars (51.4
\%). We use $AKARI$ PSC data at two bands (9 and 18 $\mu$m) and BSC
data at four bands (65, 90, 140 and 160 $\mu$m).

In Fig. 1, $AKARI$ and $IRAS$ PSC fluxes are compared for the
cross-identified objects. We plot only the objects with good
quality (q=3) data at any wavelengths. It is evident that there is
a close correlation between two sets of data.

Some objects on the left panel of Fig. 1 are too far off from the
general trend. This may be because they are large amplitude
pulsators or because they are not AGB stars (possibly post-AGB
stars). We mark the candidates for post-AGB stars (see Section 4.2)
on both panels of Fig. 1. We find that the large scatters happen
only on the left panel for the candidate stars probably because the
shapes of their SEDs are somewhat different from those for typical
AGB stars.

$IRAS$ LRS ($\lambda$ = 8$-$22 $\mu$m) data can be used for
comparison with $AKARI$ fluxes at 9 and 18 $\mu$m. We use
interpolated values of the spectral fluxes at 9 and 18 $\mu$m using
a cubic spline method. In Fig. 2, we compare the fluxes at 9 and 18
$\mu$m obtained from $IRAS$ LRS data with those from $AKARI$ data
for 922 O-rich stars and 703 C-rich stars in the sample. Dim
sources are brighter in $IRAS$ LRS than in $AKARI$ PSC. The
difference could be attributed to the difference in the spatial
resolution as Ishihara et al. (2010) pointed out to explain a
similar phenomenon. $IRAS$ LRS measures the total flux of the dim
emission because of the larger aperture size
(15$\arcmin$$\times$6$\arcmin$), whereas $AKARI$ measures the flux
of the peak emission with a smaller aperture (бн9$\arcsec$ beam
size). We find that the effect of dim sources brighter with $IRAS$
LRS data is more enhanced at 18 $\mu$m if compared with the effect
at 9 $\mu$m. This could be due to the fact that the detectable
image size at 18 $\mu$m is much larger than the one at 9 $\mu$m for
dusty AGB stars.

We may also use $ISO$ Short Wavelength Spectrometer (SWS; $\lambda$
= 2.4$-$45.4 $\mu$m) data for comparison with $AKARI$ fluxes at 9
and 18 $\mu$m. Sloan et al. (2003) presented the $ISO$ SWS catalog
which contains the high resolution spectral data for 1271
observations. We have cross-identified 121 O-rich stars (185
observations) and 68 C-rich stars (105 observations) from the $ISO$
SWS catalog by finding the closest counterpart in the position
within 30$\arcsec$. Unlike $IRAS$ LRS data, most $ISO$ SWS spectral
data show severe scatters and/or noises. Therefore, we obtain the
$ISO$ SWS fluxes at 9 and 18 $\mu$m using linear least squares
regression in the wavelength range of 0.1 $\mu$m from the central
wavelengths. We find that the results obtained by using a cubic
polynomial regression method in a wider wavelength range show only
minor differences. By examining the SEDs with the $ISO$ SWS data
and other photometric data ($IRAS$, $MSX$ and $AKARI$), we exclude
some $ISO$ SWS data that are not usable. We compare fluxes at 9 and
18 $\mu$m obtained from $ISO$ SWS data with those from $AKARI$ data
in Fig. 3. As expected, the correlations show relatively large
scatters but do not show the effect of dim sources brighter with
$ISO$ SWS data. This could be because $ISO$ SWS used a smaller
apertures size (14$\arcsec$$\times$20$\arcsec$-27$\arcsec$) than
$IRAS$ LRS.

\begin{figure}
\smallplot{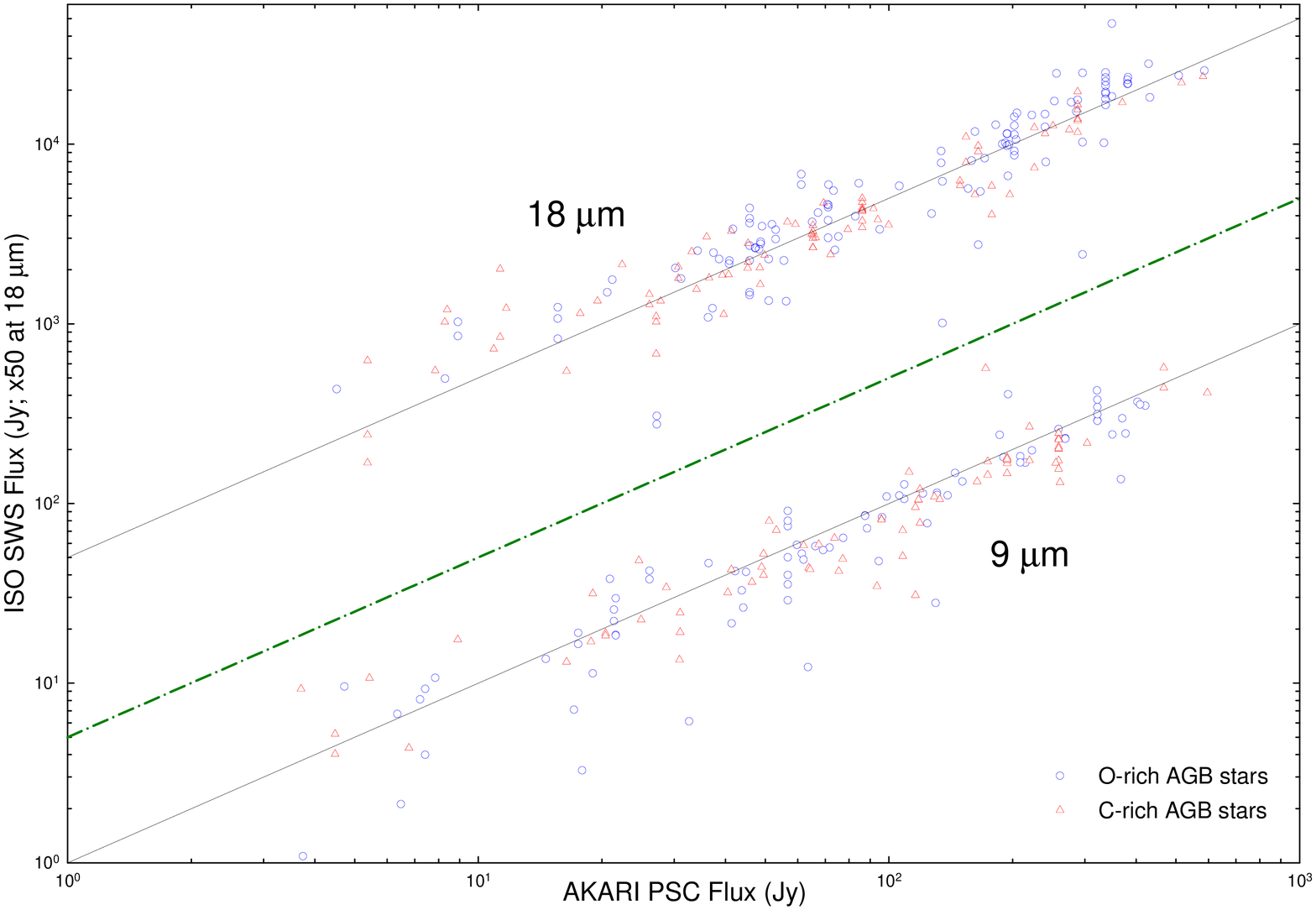} \caption{Comparison of $ISO$ SWS and $AKARI$ PSC
data for AGB stars.}
\end{figure}

\subsection{$MSX$ data}

The $MSX$ (6C) PSC (Egan et al. 2003) provided useful photometric
data at 8.28, 12.13, 14.65, 21.34 $\mu$m wavelength bands for
441,879 sources. By using the position information in the $IRAS$
PSC, we cross-identify the $MSX$ (6C) PSC source by finding the
nearest one. We have found that the closest counterparts in the
position within 10$\arcsec$ (30$\arcsec$) are 1575 (1966) O-rich
stars, 609 (687) C-rich stars, 125 (128) S stars and 19 (22)
silicate carbon stars. We have found that most objects in the
deviation range 10 - 30$\arcsec$ are correct counterparts.
Therefore, we use the closest counterparts in the position within
30$\arcsec$ for the $MSX$ (6C) PSC.

\section{Two-Colour Diagrams}

Only a relatively small number of AGB stars have complete or nearly
complete spectral energy distributions (SEDs). A large number of
stars have infrared fluxes from $IRAS$, $MSX$, $AKARI$ and NIR
data. Although less useful than a full SED, the large number of
observations with less extensive wavelength coverage can be used to
form 2CDs that can be compared to theoretical model predictions.

The colour index is defined by
\begin{eqnarray}
  M_{\lambda 1} - M_{\lambda 2} &=& 2.5 \log_{10} {{F_{\lambda 2} / ZMC_{\lambda 2}} \over {F_{\lambda 1} / ZMC_{\lambda 1}}}
\end{eqnarray}
\noindent where $ZMC_{\lambda i}$ means the zero magnitude
calibration at given wavelength ($\lambda i$). The magnitude scales
for $IRAS$, $MSX$ and $AKARI$ photometric systems are given in the
corresponding explanatory supplements ($IRAS$: Joint IRAS Science
Working Group 1986; $MSX$: Egan et al. 2003; $AKARI$ IRC: Tanabe et
al. 2008; $AKARI$ FIS: AKARI/FIS Data Reduction Support Page).
Table 5 lists the zero magnitude scales.

\begin{table}
\caption{Zero magnitude scales. \label{tbl-5}}
\begin{flushleft}
\begin{tabular}{llll}
\hline \hline
    &   Band    &   $\lambda$ ($\mu$m)  &   $ZMC_{\lambda}$ (Jy)    \\
\hline
$IRAS$  &   12  &   12  &   28.3    \\
$IRAS$  &   25  &   25  &   6.73    \\
$IRAS$  &   60  &   60  &   1.19    \\
$IRAS$  &   100 &   100 &   0.43    \\
\hline
$AKARI$ &   S9W     &   9   &   56.262  \\
$AKARI$ &   L18W    &   18  &   12.001  \\
$AKARI$ &   N60     &   65  &   0.965   \\
$AKARI$ &   WIDE-S  &   90  &   0.6276  \\
$AKARI$ &   WIDE-L  &   140 &   0.1895  \\
$AKARI$ &   N160    &   160 &   0.1487  \\
\hline
$MSX$   &   A   &   8.28    &   58.49   \\
$MSX$   &   C   &   12.13   &   26.51   \\
$MSX$   &   D   &   14.65   &   18.29   \\
$MSX$   &   E   &   21.34   &   8.8 \\
\hline \hline
\end{tabular}
\end{flushleft}
\end{table}

For the large sample of AGB stars, we present various infrared 2CDs
using $IRAS$ (PSC; 12, 25, 60 and 100 $\mu$m), $AKARI$ (PSC and
BSC; 9, 18, 65, 90, 140 and 160 $\mu$m), $MSX$ (PSC; 8.28, 12.13,
14.65 and 21.34 $\mu$m) and NIR ($K$ and $L$ bands; including
$2MASS$ data at $K_S$ band) data. For any 2CDs using $IRAS$,
$AKARI$, $2MASS$ PSC or $AKARI$ BSC data, we plot only the objects
with good quality (q=3) at any wavelength.

We will compare the observations with theoretical models on the
2CDs in Section 5. It is well known that the position of an AGB
star on a 2CD widely varies depending on the phase of pulsation
(e.g., Suh 2004). If we are able to use the colours observed and
averaged for an entire pulsation period, the observed positions on
2CDs will be more useful to be compared with the theoretical
models.

\subsection{$IRAS$, $AKARI$ and $MSX$ 2CDs}

\begin{figure*}
\plotthree{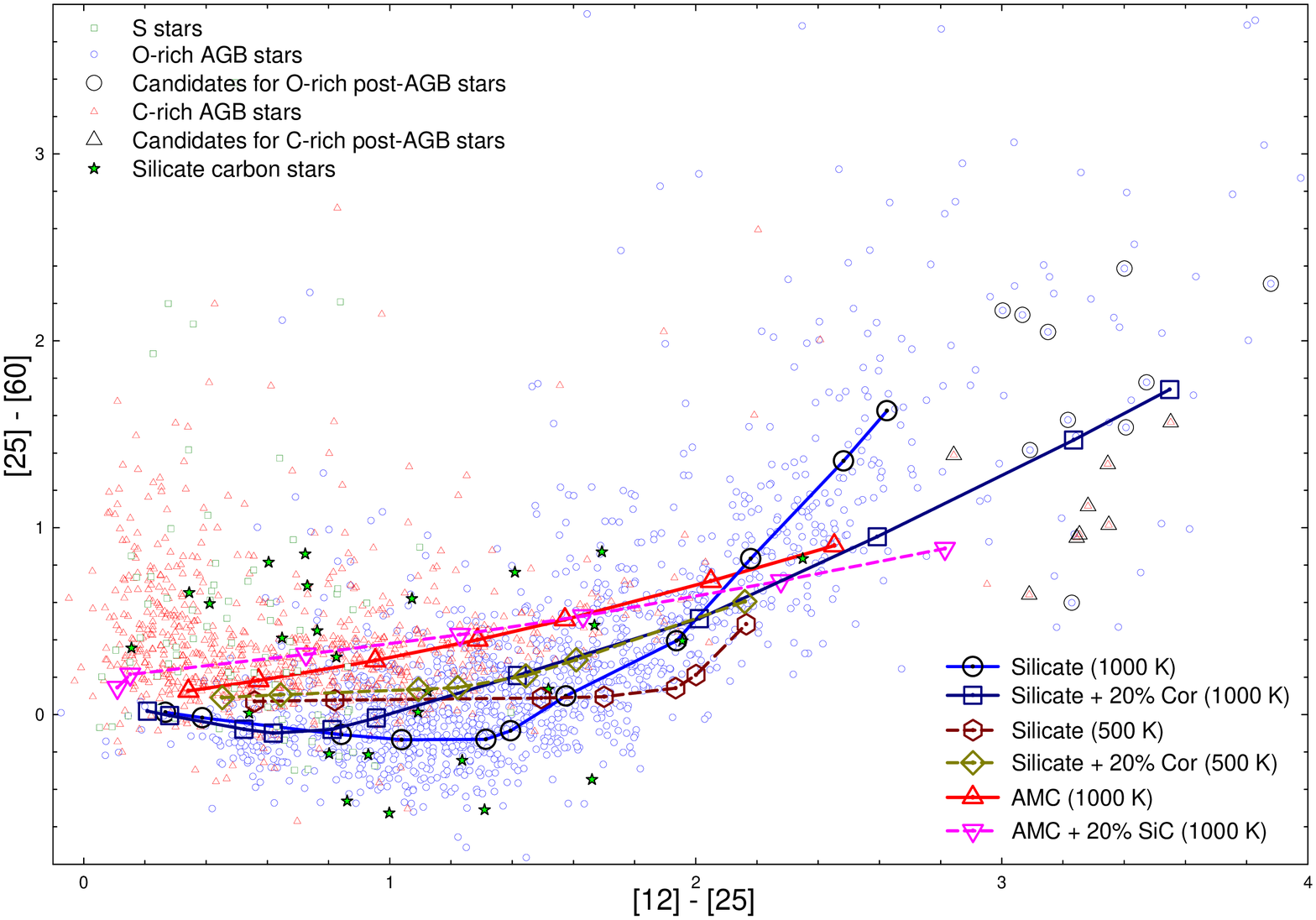}{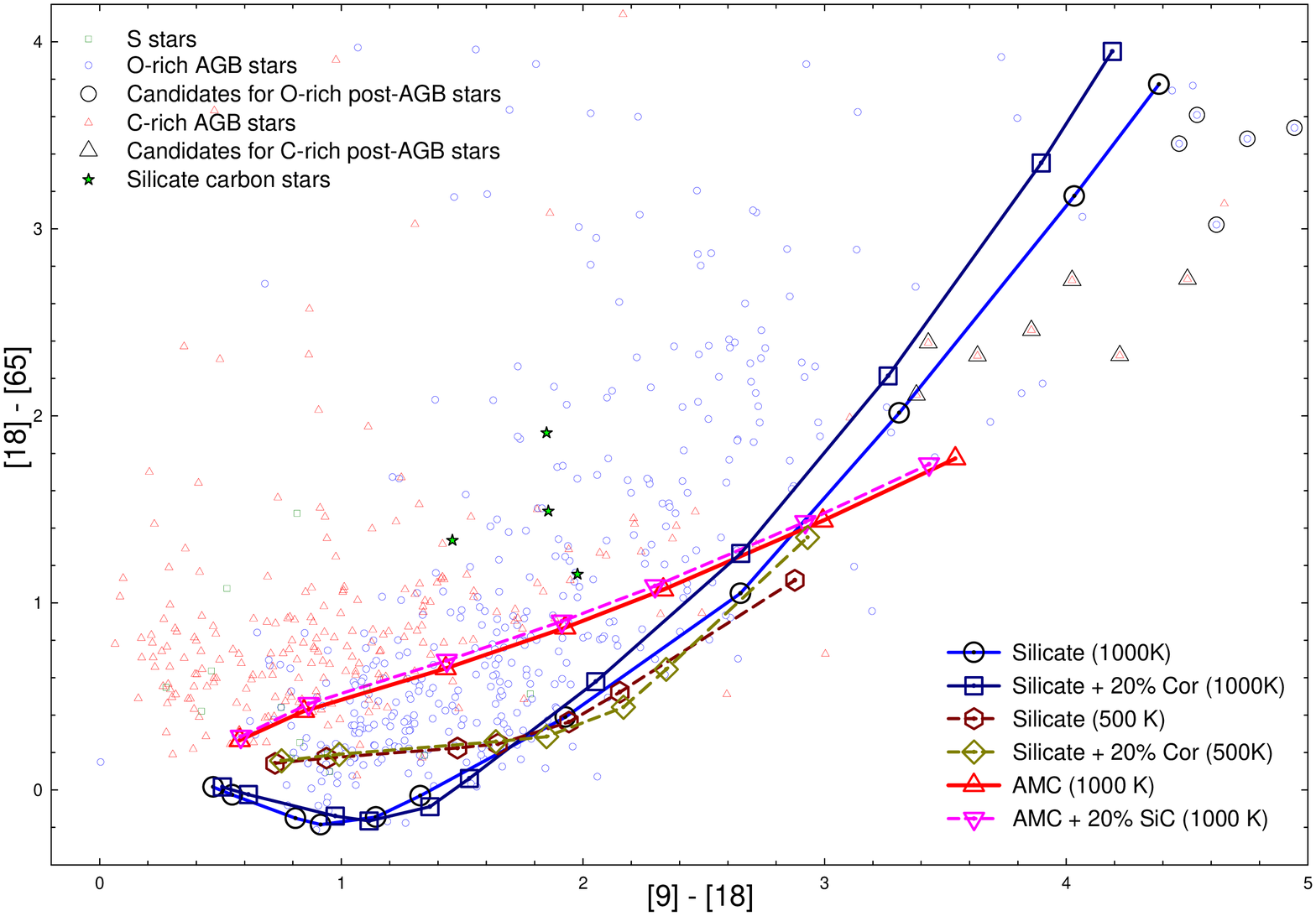}{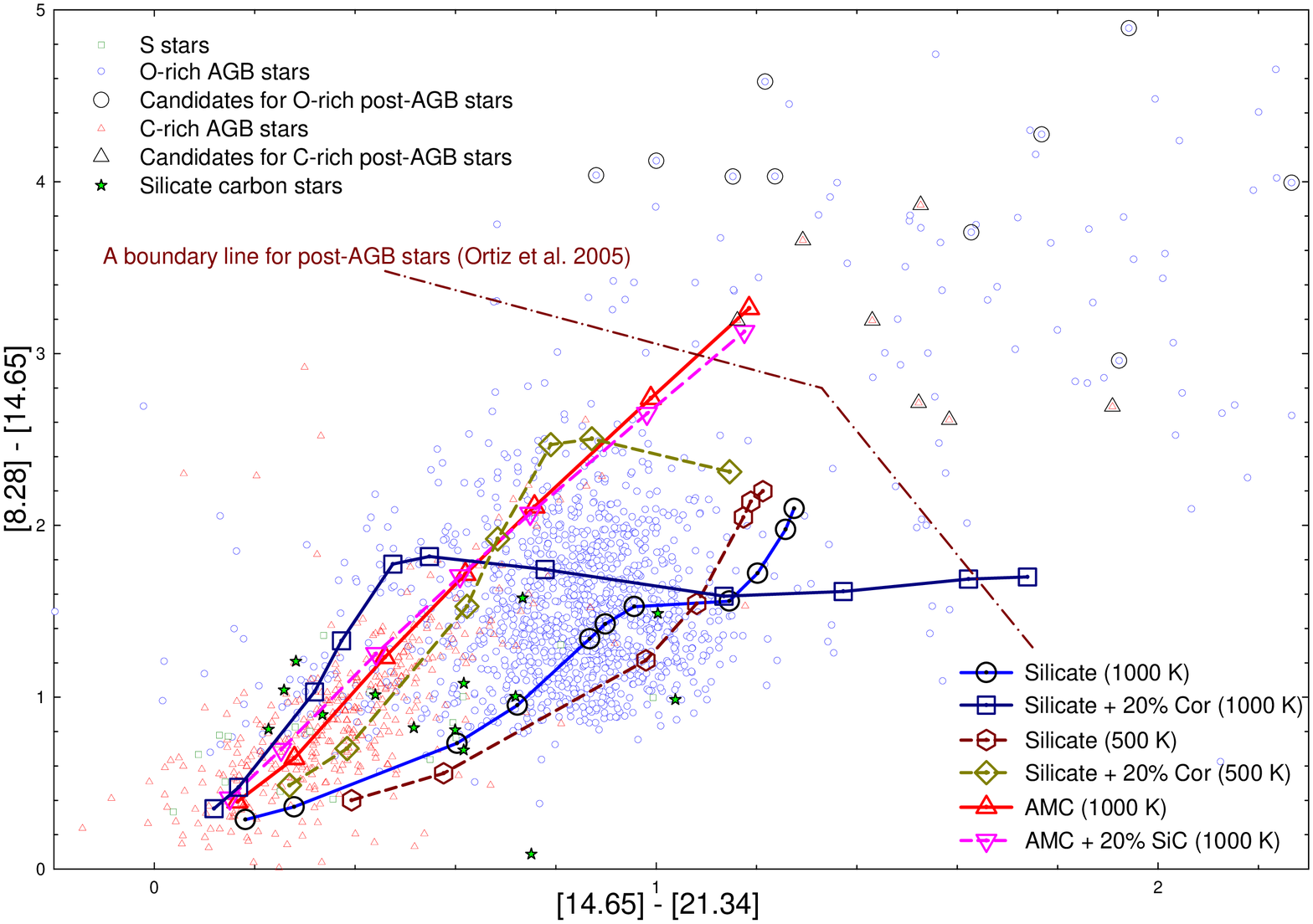} \caption{$IRAS$, $AKARI$ and
$MSX$ 2CDs for AGB stars.}
\end{figure*}

Fig. 4 shows $IRAS$, $AKARI$ and $MSX$ 2CDs for AGB stars. The
small symbols are the observational data and the lines with large
symbols are the model calculations for a range in dust shell
optical depth. We will discuss the theoretical models in Section 5.
On these diagrams, the stars in the upper-right region have thick
dust shells with large optical depths.

The upper panel of Fig. 4 plots AGB stars in an $IRAS$ 2CD using
[25]$-$[60] versus [12]$-$[25]. We find that the basic theoretical
model tracks roughly coincide with the densely populated observed
points. Because the 10 $\mu$m silicate feature changes from
emission to absorption when the dust optical depth becomes larger,
there is a change in the slope of the theoretical model line for
O-rich stars.

Carbon stars are distributed along a curve in the shape of a "C". A
group of stars in the upper left region are optical carbon stars
that show excessive flux at 60 $\mu$m which is due to the remnant
of earlier phase when the star was an O-rich AGB star (e.g., Chan
\& Kwok 1990). A group of stars in the lower region, which extend
to the right side, are infrared carbon stars. The infrared carbon
stars in the right side have thick dust envelopes with large
optical depths.

The middle panel of Fig. 4 plots AGB stars on an $AKARI$ 2CD using
[18]$-$[65] versus [9]$-$[18]. Because the number of available
observational data is much smaller than the one for an $IRAS$ 2CD,
it is difficult to make a comparison with the theoretical model
lines. However, $AKARI$ PSC data with a combination of $IRAS$,
$MSX$ and NIR data can make various meaningful 2CDs that can be
compared with theoretical models (see Figs 5, 6 and 7).

The lower panel of Fig. 4 plots AGB stars on a $MSX$ 2CD using
[8.28]$-$[14.65] versus [14.65]$-$[21.34]. Ortiz et al. (2005) used
this 2CD for investigating the tracks of AGB to post-AGB evolution.
The boundary line for post-AGB stars as suggested by Ortiz et al.
(2005) is marked on the 2CD and we find that all the candidates for
post-AGB stars (see Section 4.2) are in the region of post-AGB
stars.

\begin{figure*}
\plottwo{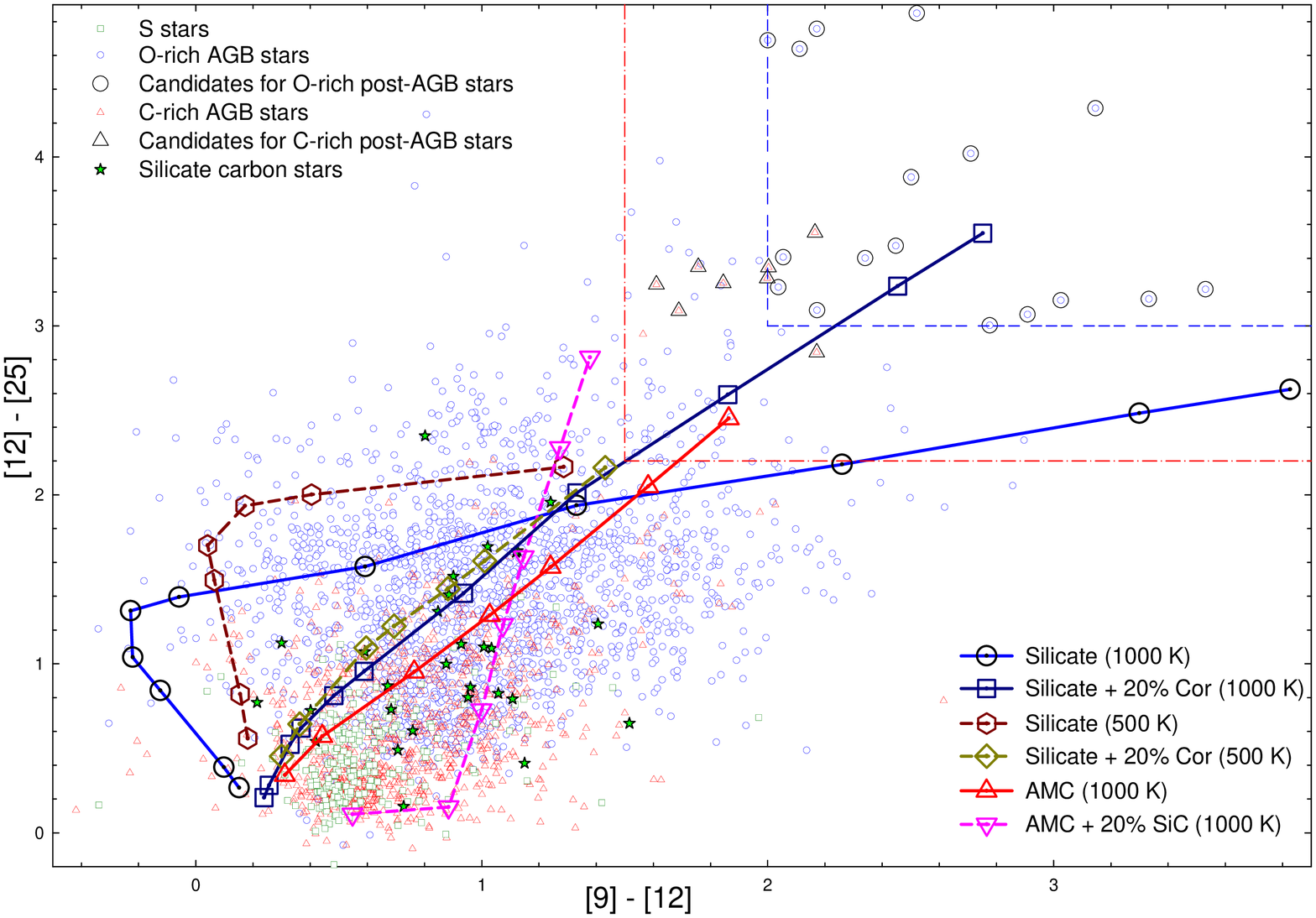}{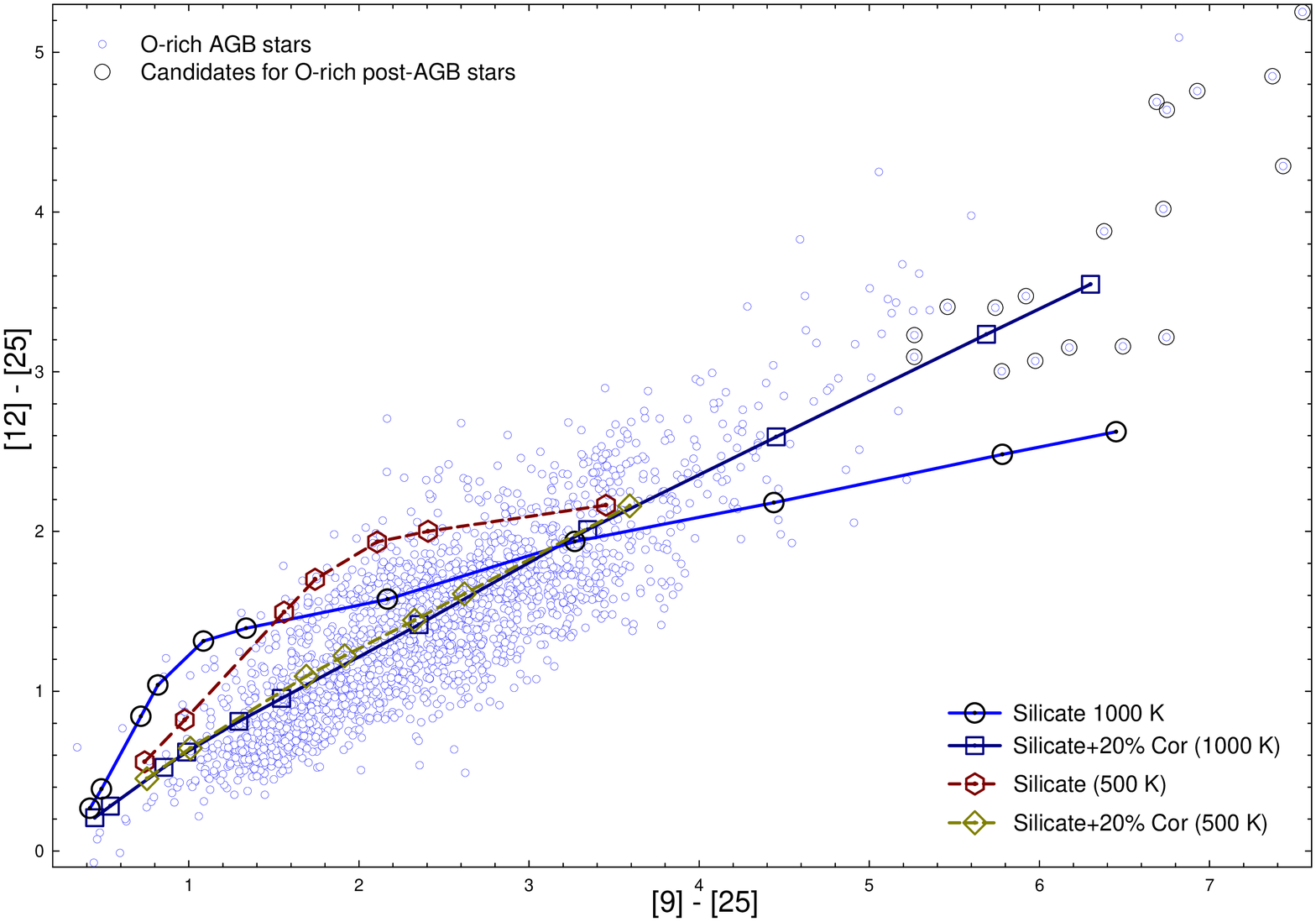} \caption{$AKARI$-$IRAS$ 2CDs for AGB
stars.}
\end{figure*}

Fig. 5 plots AGB stars on $AKARI$-$IRAS$ 2CDs. The upper panel of
Fig. 5 plots a 2CD using [12]$-$[25] versus [9]$-$[12]. The lower
panel of Fig. 5 plots a 2CD only for O-rich stars using [12]$-$[25]
versus [9]$-$[25].

\subsection{Candidates for post-AGB stars}

We may distinguish post-AGB stars from AGB stars by analyzing the
SEDs. Post-AGB stars would show generally larger [9]$-$[12] or
[12]$-$[25] colours than AGB stars (e.g., Garc\'ia-Lario et al.
1997; Sevenster 2002; Ortiz et al. 2005; Bains et al. 2009).

For simplicity, we choose the objects with large [9]$-$[12] and
[12]$-$[25] colours as the candidates for post-AGB stars. For
O-rich stars, we choose the sources with [9]$-$[12] $\geq$ 2 and
[12]$-$[25] $\geq$ 3. For C-rich stars, we choose the ones with
[9]$-$[12] $\geq$ 1.5 and [12]$-$[25] $\geq$ 2.2. We choose these
boundaries because most AGB stars are populated inside. We find
that 18 O-rich stars and 8 C-rich stars in our sample satisfy the
condition (see the upper panel of Fig. 5). They are likely to be
post-AGB stars and are listed in Table 6.

The candidates for post-AGB stars are marked in all 2CDs (in Figs
4, 5, 6 and 7) and in Fig. 1. Some stars in an outer region on one
2CD can be located in a populated AGB region on another 2CD.
However, the positions of all the candidate stars locate in outer
regions on all the 2CDs presented in this paper implying that they
are good candidates. Note that the number of the objects for each
2CD can be different depending on availability of the data. For
example, three objects were not observed at $K$ band as indicated
in Table 6.

\begin{table}
\caption{Candidates for post-AGB stars. \label{tbl-6}}
\begin{flushleft}
\begin{tabular}{llll}
\hline \hline
Object &   [9-12]     &[12-25]  &        Note \\
($IRAS$ PSC \#) & &  &   \\
\hline
16115-5044  &   2.00    &   4.69    &   O-rich  \\
16342-3814  &   3.15    &   4.29    &   O-rich  \\
17168-3736  &   2.78    &   3.00    &   O-rich  \\
17385-3332  &   3.53    &   3.22    &   O-rich  \\
17436+5003  &   2.30    &   5.25    &   O-rich  \\
17499-3520  &   2.04    &   3.23    &   O-rich  \\
17560-2027  &   2.71    &   4.02    &   O-rich  \\
18051-2415  &   3.02    &   3.15    &   O-rich  \\
18087-1440  &   2.50    &   3.88    &   O-rich (no $K$) \\
18135-1456  &   2.91    &   3.07    &   O-rich  \\
18276-1431  &   2.45    &   3.47    &   O-rich  \\
18450-0148  &   3.33    &   3.16    &   O-rich  \\
18451-0332  &   2.17    &   4.76    &   O-rich  \\
18518+0558  &   2.17    &   3.09    &   O-rich (no $K$) \\
18596+0315  &   2.34    &   3.40    &   O-rich  \\
19024+0044  &   2.11    &   4.64    &   O-rich  \\
19114+0002  &   2.52    &   4.85    &   O-rich  \\
20406+2953  &   2.05    &   3.41    &   O-rich  \\
\hline
07134+1005  &   1.84    &   3.25    &   C-rich  \\
19454+2920  &   2.00    &   3.35    &   C-rich  \\
19477+2401  &   2.00    &   3.28    &   C-rich  \\
19480+2504  &   2.17    &   2.84    &   C-rich  \\
20000+3239  &   1.61    &   3.24    &   C-rich  \\
22272+5435  &   1.69    &   3.09    &   C-rich  \\
23304+6147  &   1.76    &   3.35    &   C-rich  \\
23321+6545  &   2.17    &   3.55    &   C-rich (no $K$) \\
\hline \hline
\end{tabular}
\end{flushleft}
\end{table}

\subsection{$AKARI$-$IRAS$-NIR 2CDs}

\begin{figure*}
\plottwo{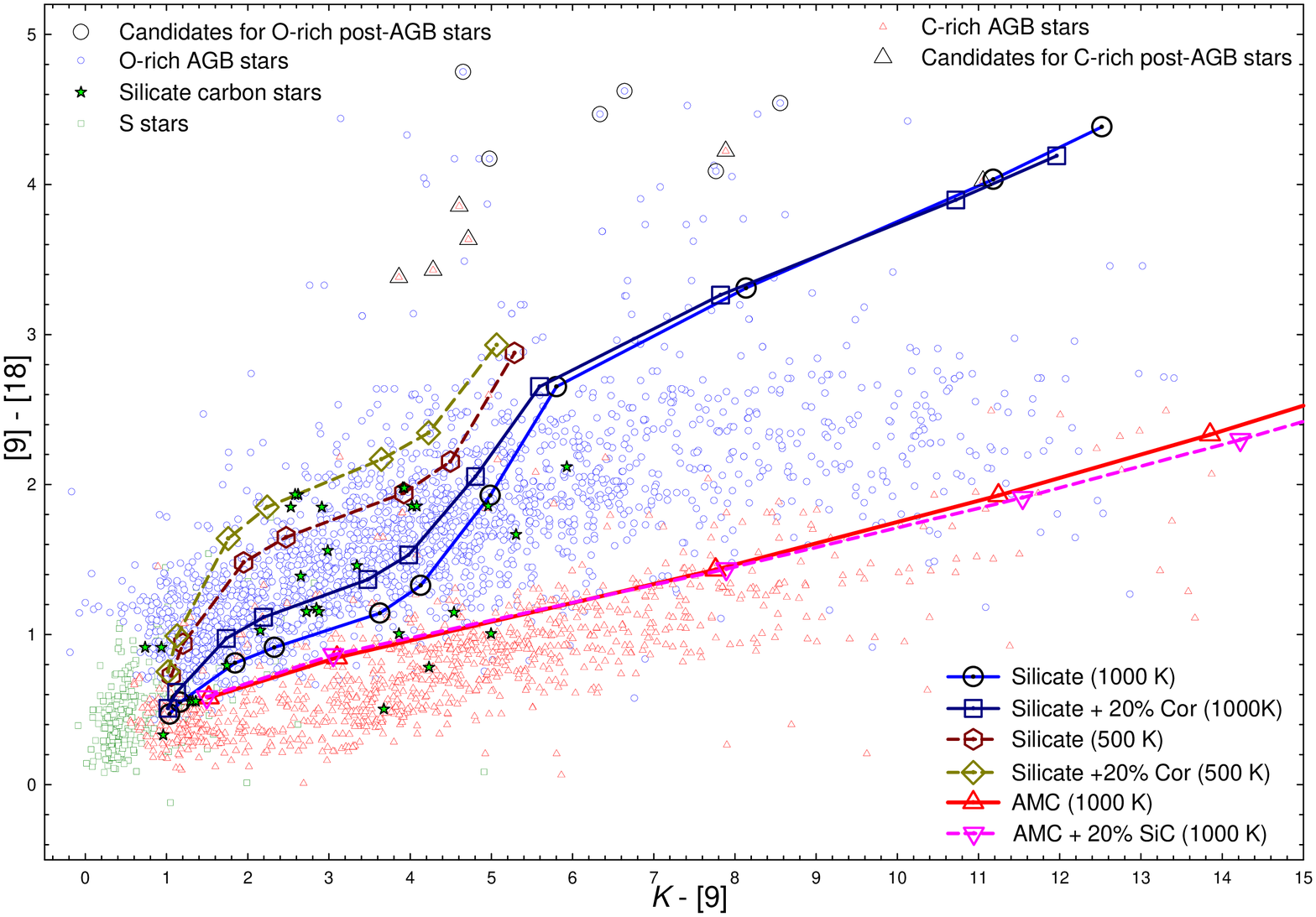}{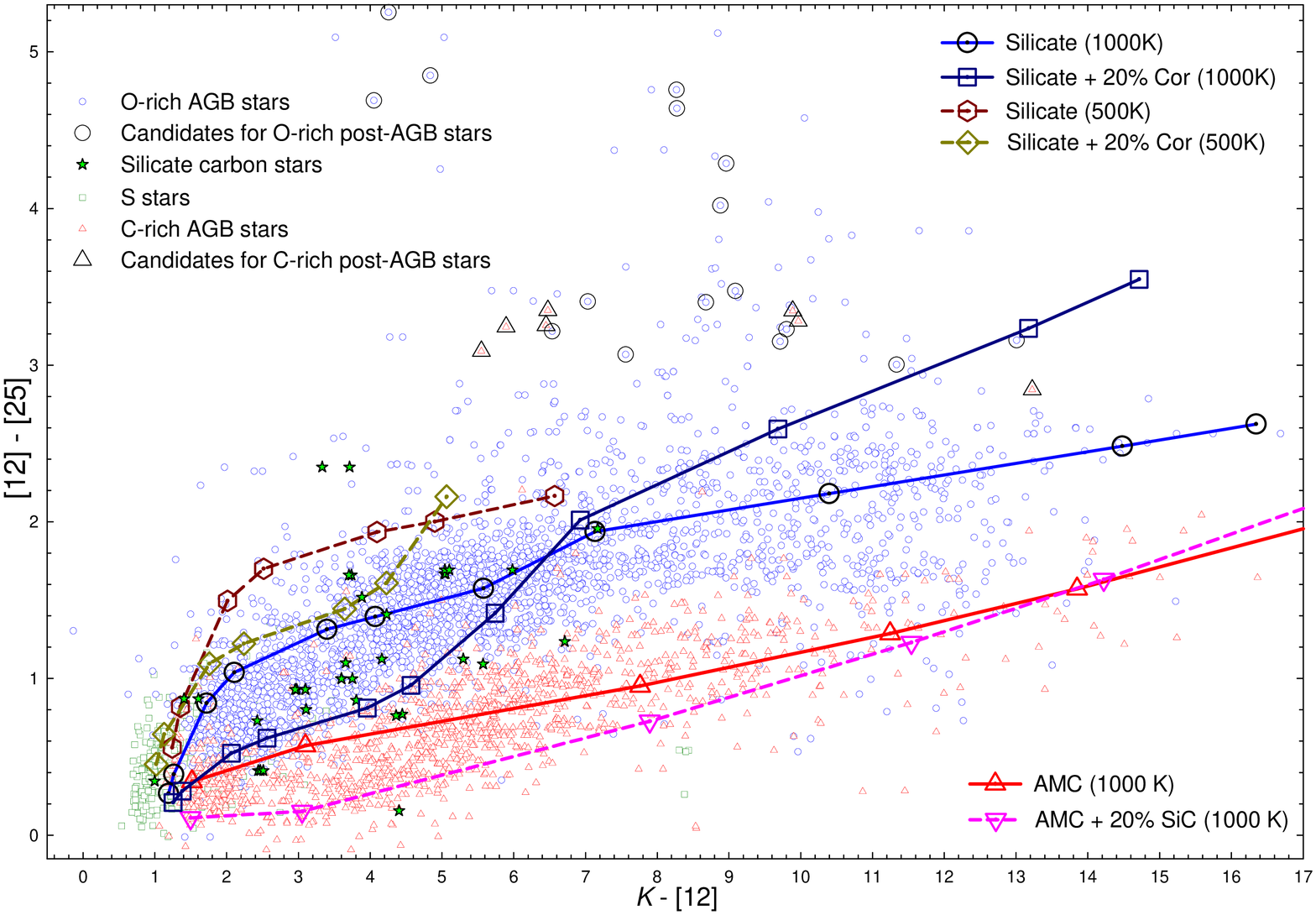} \caption{$AKARI$-$IRAS$-NIR (the $K$ band)
2CDs for AGB stars.}
\end{figure*}

\begin{figure*}
\plottwo{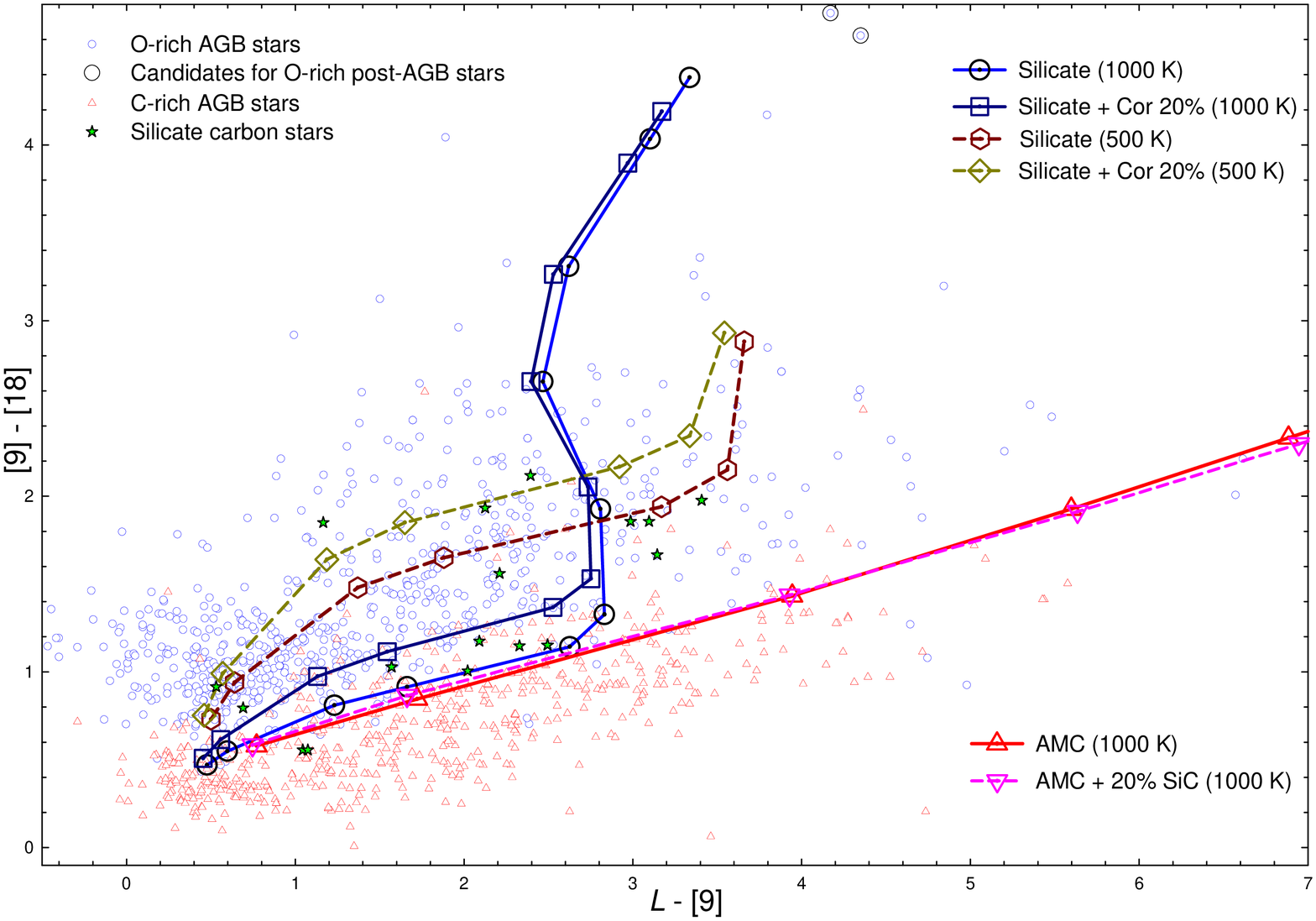}{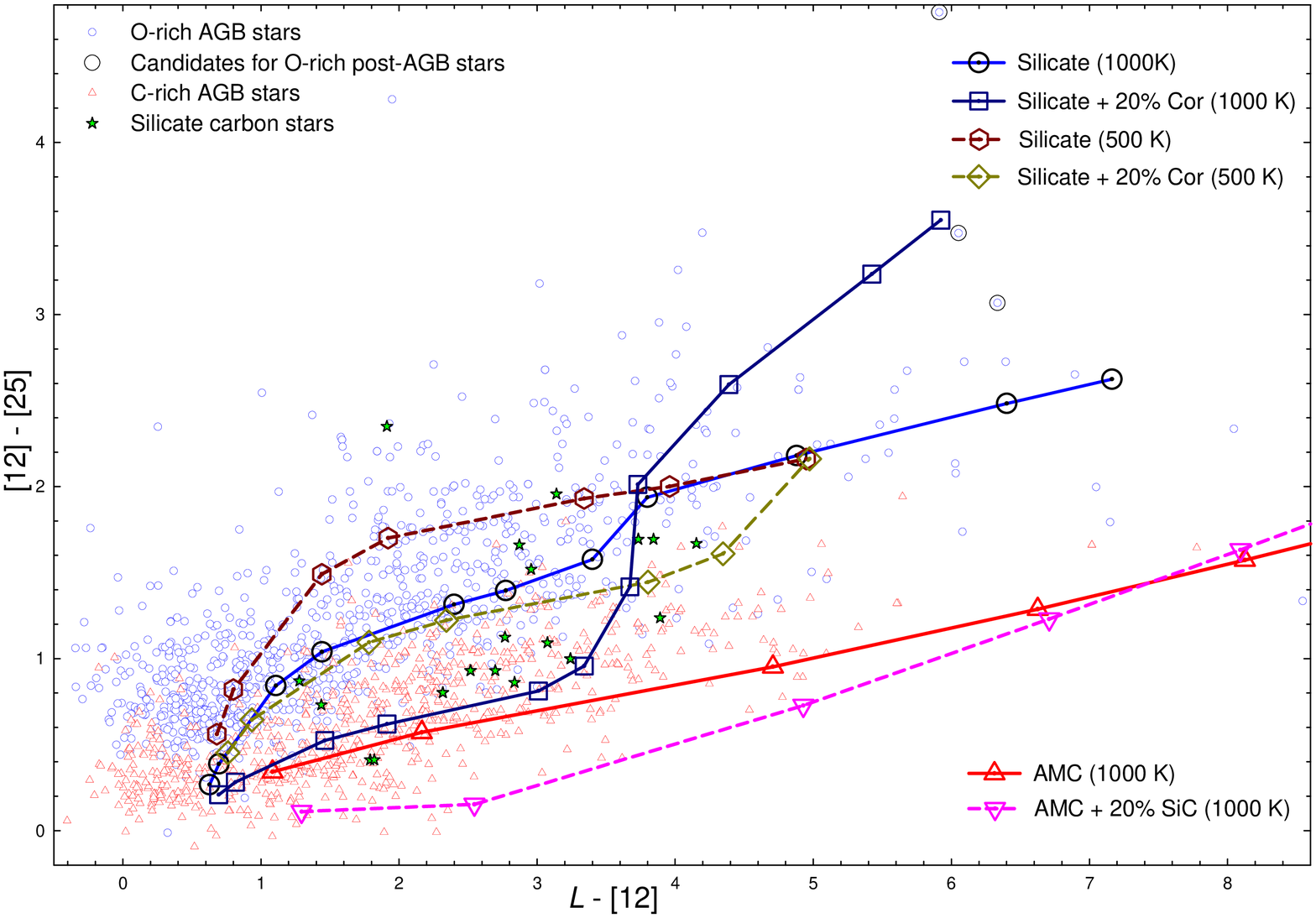} \caption{$AKARI$-$IRAS$-NIR (the $L$ band)
2CDs for AGB stars.}
\end{figure*}

Figs 6 and 7 show 2CDs using $AKARI$, $IRAS$ and NIR data ($K$ and
$L$ bands). If we compare them with Figs 4 and 5, we find that the
boundaries which separate O-rich, C-rich and S stars are more clear
in $AKARI$-$IRAS$-NIR 2CDs.

Fig. 6 displays two 2CDs using $K$ band data (the [9]$-$[18] versus
$K-$[9] and [12]$-$[25] versus $K-$[12] plots). We find that the
basic theoretical model tracks roughly coincide with the densely
populated observed points.

Fig. 7 shows two 2CDs using $L$ band data (the [9]$-$[18] versus
$L-$[9] and [12]$-$[25] versus $L-$[12] plots). In our catalog, the
number of available NIR data at $L$ band is much smaller than the
number at $K$ band. The observed colours generally fit the
theoretical model lines.

\section{Theoretical Models}

For this paper, we use the radiative transfer code DUSTY developed
by Ivezi\'{c} \& Elitzur (1997) for a spherically symmetric dust
shell. We have performed the model calculations in the wavelength
range 0.01 to 36000 $\mu$m.

For all the models, we assume the dust density distribution is
inversely proportional to the square of the distance ($\rho \propto
r^{-2}$). The dust condensation temperature ($T_c$) is assumed to
be 1000 K and 500 K. The outer radius of the dust shell is always
taken to be $10^4$ times the inner radius ($R_c$).

\subsection{O-rich stars}

For O-rich stars, we use a simple mixture of silicate and corundum
(20 \% by mass) dust grains as well as pure silicate. For silicate,
we use the optical constants of warm and cold silicate grains
derived by Suh (1999). Corundum (Cor) grains are thought to be the
earliest products of the dust condensation process around O-rich
AGB stars (e.g., Maldoni et al. 2005). We use the optical constants
of amorphous corundum (Al$_2$O$_3$; porous) obtained by Begemann et
al. (1997).

The radii of spherical dust grains have been assumed to be 0.1
$\mu$m uniformly. We choose 10 $\mu$m as the fiducial wavelength
that sets the scale of the optical depth ($\tau_{10}$) and compute
models for eleven optical depths ($\tau_{10}$ $=$ 0.005, 0.01,
0.05, 0.1, 0.5, 1, 3, 7, 15, 30 and 40). For the central star, we
assume that the luminosity is $10^4$ $L_{\odot}$ and the stellar
blackbody temperature is 2500 K for $\tau_{10}$ $\leq 3$ and 2000 K
for $\tau_{10}$ $> 3$. Also, we use the warm silicate dust grains
for LMOA stars (7 models with $\tau_{10}$ $\leq 3$) and the cold
grains for HMOA stars (4 models $\tau_{10}$ $> 3$).

The track line of the theoretical model results for silicate dust
with $T_c$ = 1000 K (TLS) on a 2CD can show various interesting
behaviors because the 10 $\mu$m silicate feature changes from
emission to absorption when the dust optical depth becomes larger
($\tau_{10}$ $> 3$). The TLS for increasing optical depths show
wide variations in the inclination (see Figs 4, 5, 6 and 7)
including backward effects (see Figs 5 and 7).

On the upper panel of Fig. 5, the TLS in the horizontal direction
([9]$-$[12] colour) initially moves backward when the 10 $\mu$m
silicate emission feature becomes more prominent then moves forward
when the 10 $\mu$m silicate feature changes from emission to
absorption.

On the upper panel of Fig. 7, the TLS in the horizontal direction
($L-$[9] colour) initially moves forward rapidly when the 10 $\mu$m
silicate emission feature becomes more prominent. Then it moves
backward when the 10 $\mu$m silicate feature changes from emission
to absorption and then moves forward very slowly in the horizontal
direction. This effect becomes much weaker for the 2CD using the
$L-$[12] colour (the lower panel of Fig. 7) because the influence
of the 10 $\mu$m silicate feature at 12 $\mu$m is weaker than at 9
$\mu$m. The backward effect does not happen for the $K-$[9] colour
(see the upper panel of Fig. 6) because the 10 $\mu$m silicate
feature is more distant from the $K$ band.

Suh (2004) pointed out that a low dust condensation temperature is
generally required for LMOA stars with thin dust envelopes. The
models with a low dust condensation temperature ($T_c$ = 500 K) for
seven models of LMOA stars with thin dust envelopes ($\tau_{10}$
$=$ 0.005, 0.01, 0.05, 0.1, 0.5, 1 and 3) are also shown for
comparison. It is clear that the locations of observed LMOA stars
on 2CDs are close to the theoretical model lines with a low dust
condensation temperature (See Figs 4, 5, 6 and 7). The models with
a low dust condensation temperature ($T_c$ = 500 K) can make a
better fit with observations of LMOA stars. When we use different
dust condensation temperatures from 500 to 1000 K, we find that a
series of model tracks with lower temperatures (e.g., 500-700 K)
can cover an area of LMOA stars fairly well.

Amorphous corundum (Al$_2$O$_3$; Begemann et al. 1997) grains
produce a single peak at 11.8 $\mu$m and influences the shape of
the SED at around 10 $\mu$m. The shape of the 10 $\mu$m feature of
O-rich AGB stars, which is mainly produced by silicate, can be
modified by addition of corundum dust. The $AKARI$, $IRAS$ and
$MSX$ flux data at 12, 9, 14.65 and 8.28 $\mu$m would exhibit the
presence of corundum dust efficiently.

Many 2CDs displayed in Figs 4, 5, 6 and 7 show that a mixture of
silicate (80 \% by mass) and corundum (20 \%) can make a better fit
with observations of O-rich stars rather than pure silicate. When
we change the abundance of corundum from 0 to 20 \% gradually, we
find that the series of model tracks can cover a wide area of the
most populated observation points fairly well for many 2CDs. This
effect is more clear on the 2CD using [12]$-$[25] versus $K-$[12]
(the lower panel of Fig. 6). Other 2CDs using the fluxes at 12 and
9 $\mu$m also show the effect well. The effect of corundum dust
would be prominent for the 2CDs using the fluxes at 12, 9, 14.65
and 8.28 $\mu$m. However, the effect of corundum abundance may not
improve the fitting with observations for some 2CDs (e.g., the 2CD
in the lower panel of Fig. 7).

\subsection{C-rich stars}

For C-rich stars, we use the optical constants of amorphous carbon
(AMC) grains derived by Suh (2000) and the optical constants of
$\alpha$ SiC grains by P\'{e}gouri\'{e} (1988). The radii of
spherical dust grains have been assumed to be 0.1 $\mu$m uniformly.
We choose 10 $\mu$m as the fiducial wavelength that sets the scale
of the optical depth($\tau_{10}$) and perform the model
calculations for seven optical depths ($\tau_{10}$ $=$ 0.01, 0.1,
1, 2, 3, 5 and 7). For the central star, we assume that the
luminosity is $10^4$ $L_{\odot}$ and the stellar blackbody
temperature is 2300 K for $\tau_{10}$ $\leq 0.1$ and 2000 K for
$\tau_{10}$ $> 0.1$. For C-rich stars, we use a simple mixture of
AMC and SiC (10 \% by mass) dust grains.

It is known that a portion (10 - 20 \%) of SiC dust grains produce
the prominent SiC emission features at 11.3 $\mu$m for C-rich AGB
stars (e.g., Suh 2000). The theoretical model lines of carbon stars
on $AKARI$-$IRAS$-NIR 2CDs show that the effect of SiC is only
minor at 9 $\mu$m (see upper panels of Figs 6 and 7). On the other
hand, the effect of SiC is evident on the 2CDs using 12 $\mu$m
fluxes (see lower panels of Figs 6 and 7). The most prominent
effect of SiC can be seen on the 2CD using [12]$-$[25] versus
[9]$-$[12] (the upper panel of Fig. 5). By changing the abundance
of SiC from 0 to 20 \%, we find that the model tracks can cover the
area of the well populated observation points.

\subsection{Discussion on dust opacity}

The $AKARI$ and the $MSX$ provided good quality photometric data
for a large sample of AGB stars at wavelengths (9, 18, 65, 90, 140
and 160; 8.28, 12.13, 14.65 and 21.34 $\mu$m) not observed before.
Therefore, the new 2CDs using $AKARI$, $IRAS$, $MSX$ and NIR data
may provide new opportunities to further understand about the dust
around AGB stars. Even though we could make more or less proper
explanations for the observations on the various 2CDs using the
theoretical models, we may need to improve the theoretical models
in the light of the new observations.

Possibly due to three reasons, it is difficult to find the
theoretical models which make better fits with the observations on
the various 2CDs in a consistent way. First, we may have not
considered some important dust species. Secondly, the dust opacity
functions used for this work may need to be improved. Finally, the
radiative transfer model could be too simple.

For O-rich stars, the models with pure silicate dust opacity do not
make satisfactory fits for some 2CDs using $AKARI$, $IRAS$ and
$MSX$ data. As we discussed in Section 5.1, this could be due to
presence of corundum dust. Other important dust species such as
MgFeO series may need to be considered for O-rich stars. The dust
grains of MgFeO series (Henning et al. 1995) produce single peaks
at 15-22 $\mu$m. They could be responsible for $AKARI$ and $MSX$
fluxes in the 15-22 $\mu$m region.

The opacity functions for amorphous silicate and AMC dust grains
may need to be improved. The opacity functions, which were derived
by Suh (1999, 2000) from 2CDs and SEDs of AGB stars using the
$IRAS$ and other available data, make less satisfactory fits for
some 2CDs using $AKARI$ or $MSX$ data. The opacity in a given
wavelength region approximately matches the curve of a simple power
law ($Q_{abs} \propto \lambda^{-\beta}$) with a different spectral
index ($\beta$) in each wavelength region. More refined procedures
to obtain new spectral indexes would be useful to find the opacity
functions of silicate and AMC which make better fits with the new
2CDs.

The radiative transfer code DUSTY (Ivezi\'{c} \& Elitzur 1997)
which is used for this work can treat only a single component dust
shell. In reality, there would be multiple dust components with
different density and temperature distributions in AGB stars. For
O-rich AGB stars, it would be more reasonable to consider multiple
components of corundum and silicate dust grains rather than a
single component of a simple mixture.

Further investigations on the theoretical models compared with the
full SEDs as well as 2CDs for a large sample of AGB stars would be
necessary. We expect that a more sophisticated radiative transfer
model (e.g., a model for multiple dust components) using a more
reasonable set of various dust opacity functions would be useful.
Recently, laboratory measurements for dust in various sample shapes
(e.g., Mutschke, Min \& Tamanai 2009; Koike et al. 2010) and
environments (e.g., olivine at different temperatures: Koike et al.
2006) provided new opportunities to identify appropriate dust
species from the observations.

We did not consider opacities of molecules for radiative transfer
models in this paper. However, molecules as well as dust in the
envelopes around AGB stars contribute to the opacity (e.g.,
Cristallo et al. 2007). In NIR regions, the molecular contributions
to the observed SEDs of AGB stars are significant especially for
the AGB stars with thin dust envelopes (e.g., Suh \& Kwon 2009).
More sophisticated radiative transfer models that can treat both
molecules and dust would be necessary.

\section{Conclusions}

Using a revised version of the catalog of AGB stars by Suh \& Kwon
(2009), we have compiled the available observed fluxes at various
infrared wavelengths for 3003 O-rich, 1168 C-rich, 362 S-type and
35 silicate carbon stars in our Galaxy. For each object in the new
catalog, we have cross-identified the $AKARI$, $MSX$ and $2MASS$
counterparts by finding the nearest source using the position
information in version 2.1 of the $IRAS$ PSC.

For the large sample of AGB stars, we have presented various
infrared 2CDs using $IRAS$ (PSC), $AKARI$ (PSC and BSC), $MSX$
(PSC) and NIR ($K$ and $L$ bands; including $2MASS$ data at $K_S$
band) data for different classes of AGB stars based on the
chemistry of the dust shell and/or the central star. Using
theoretical radiative transfer models for dust envelopes around
O-rich and C-rich AGB stars, we have plotted tracks of models
results with increasing dust shell optical depths on the 2CDs.

Comparing the observations with the models on the 2CDs, we have
found that the basic model tracks roughly coincide with the densely
populated observed points. The theoretical models using the opacity
functions of amorphous silicate, AMC, SiC and corundum can explain
the observations of O-rich and C-rich AGB stars on the various 2CDs
fairly well in general. We could explain various behaviors of model
tracks on the 2CDs using the known features of observations (e.g.,
silicate and SiC features). For O-rich AGB stars, we have found
that the models using corundum dust as well as silicate can improve
the fit with the observations.

To better explain the observations of AGB stars, further
investigations using a more sophisticated radiative transfer model
with an appropriate choice of improved opacity functions for more
various dust species would be necessary.

We expect that our work would be helpful for further investigations
about AGB stars. The updated catalog data will be open to the
public through the first author's world wide web site
http://web.chungbuk.ac.kr/$\sim$kwsuh/.

\section*{Acknowledgments}
This work was supported by the Korea Science and Engineering
Foundation (KOSEF) grant funded by the Korea government (MEST)
(No.R01-2008-000-20002-0). This research has made use of the SIMBAD
database, operated at CDS, Strasbourg, France. This research is
based on observations with AKARI, a JAXA project with the
participation of ESA.

\bsp

\label{lastpage}

\end{document}